\documentclass[aps,prb,amsmath,amssymb,a4paper,reprint,tightenlines,superscriptaddress]{revtex4-2}

\usepackage{graphicx}
\usepackage[usenames,dvipsnames,svgnames]{pstricks}
\usepackage{soul}
\usepackage{stix2}

\usepackage[%
unicode=true,%
colorlinks=true,%
linkcolor=blue,anchorcolor=blue,%
breaklinks=true,%
citecolor=blue,filecolor=cyan,%
menucolor=blue,%
breaklinks=true,%
urlcolor=cyan]{hyperref}

\newcommand{\be}{\begin{equation}}
\newcommand{\ee}{\end{equation}}
\newcommand{\e}{\mathrm{e}}

\newcommand{\md}{\mathrm{d}}
\newcommand{\nn}{\nonumber}
\newcommand{\eF}{\epsilon_\mathrm{_F}}
\newcommand{\Fref}[1]{Fig.~\ref{#1}}
\newcommand{\Eqref}[1]{Eq.~(\ref{#1})}

\begin{document}

\title{Kondo \textit{s-d} exchange in the CuO\textsubscript{2} plane
as the long-sought interaction determining
$\boldsymbol{T_c}$ in cuprates}

\author{Todor M. Mishonov}
\email{mishonov@gmail.com}
\affiliation{Institute of Solid State Physics, Bulgarian Academy of
Sciences, 72 Tzarigradsko Chauss\'ee, 1784 Sofia, Bulgaria}
\author{Nedeltcho I. Zahariev}
\email{nedeltcho.zahariev@gmail.com}
\affiliation{Institute of Solid State Physics,    
Bulgarian Academy of Sciences, 
72 Tzarigradsko Chauss\'ee, 1784 Sofia, Bulgaria}

\author{Hassan Chamati}
\email{chamati@issp.bas.bg}
\affiliation{Institute of Solid State Physics, Bulgarian Academy of Sciences, 
72 Tzarigradsko Chauss\'ee, 1784 Sofia, Bulgaria}
\affiliation{National Centre of Excellence in Mechatronics and Clean
Technologies, 8 Kliment Ohridski Blvd., Blk. 8, 1756 Sofia, Bulgaria}
%    
%   \author{Joseph O. Indekeu}
%   \email{joseph.indekeu@kuleuven.be}
%   \affiliation{Institute for Theoretical Physics, 
%   KU Leuven,   Celestijnenlaan 200d - box 2415,
%   BE-3001 Leuven, Belgium}

\author{Albert M. Varonov}
\email{varonov@issp.bas.bg}
\affiliation{Institute of Solid State Physics,
    Bulgarian Academy of Sciences, 
   72 Tzarigradsko Chauss\'ee, 1784 Sofia, Bulgaria}
\date{\today}
%\date{
% 2023: Feb 17\\
% 2024:
% Jan 13, 20,
% Feb 3, 10,
% Mar 2,
% May 1, 26,
% Jun 3-7,
% Jul 26-31,
% Aug 3-4, 13
% Sep 21-24
% Nov 4; New text section added at the end of the manuscript
% Nov 5; Zeta function of an operator
% Nov 26; Compilation only
% Nov 28; New texts in color in the main text
% Dec 1: Submission tho Physica C
% Dec 18, 20 and 21: Reply to criticism
% Albert can find the suggested by the reviewers reference and that is it.
% Dec 29, 2:34 Tod is dotuk.
% Jan 12, second round; 19 Letter to editor; 20 small elaborations 
% Jan 20, some suggested new references should be found and red
% Jan 21, Tod in ISSP, chatty homespun
% Jan 26. Sunday Alb and Tod
% Jan 27, bliso do kraya na dosadata.
% Jan 30, pred krajna versija 
% Feb 2, to be finalized; 4 homework, 5 saturation and stop.
% March 9, Physica B
% March 23
% April 27 frozen 
%}

\begin{abstract}
The well-known Pavarini \textit{et al.}~[\href{https://doi.org/10.1103/PhysRevLett.87.047003}{Phys. Rev. Lett. \textbf{87}, 047003 (2001)}] 
correlation between the critical temperature $T_{c,\,\mathrm{max}}$
and the shape of the Fermi contour of optimally hole-doped
cuprates is explained within the framework of the BCS theory
with Kondo exchange interaction serving as a pairing mechanism.
The strong influence of the relative position of the Cu4$s$ level 
with respect to the Cu3$d_{x^2-y^2}$ level on the critical temperature
$T_c$ reveals the importance of the $s$-$d$ hybridization of the conduction band.
This hybridization is proportional to the $s$-$d$ exchange scattering amplitude between the conduction electrons -- the mechanism of $d$-wave pairing in the CuO$_2$ plane.
In other words, the Kondo interaction considered as a pairing mechanism in the CuO$_2$ plane provides a natural explanation of the correlation between the critical temperature and the shape of the Fermi contour.
This result suggests that the long-sought pairing mechanism in
high-$T_c$ superconducting cuprates has possibly been found and that the Kondo exchange interaction as a property of strongly correlated quantum matter deserves further attention in the physics of layered cuprates.
To test the developed theoretical scheme,
we explored in detail the recent results of the scanned Josephson modulation microscopy
experiment of the modulation of the order parameter due to 
apex distance super-modulation. Our analysis shows a satisfactory
agreement between our theory and the named experiment.
\end{abstract}

\maketitle
\clearpage

\section{Introduction. What determines $T_{c,\,\mathrm{max}}$?}

Condensed matter is undoubtedly one of the most sophisticated
fields of physics, if not science.
One of its most important and still open problems is the mechanism of
high-$T_c$ superconductivity (HTS) discovered more than 30 years ago \cite{Bednorz:86}.
On the problem of models and mechanisms of the HTS phenomenon
several conferences were held \cite{conferences} and a review of the suggested ideas can be easily formulated:
\textit{almost all processes in condensed matter physics have been
considered as a possible mechanism of HTS.}
%In the heroic initial period of HTS physics, up to 1/8th of the
%submitted manuscripts to Physical Review Letters were devoted to the
%HTS problem~\cite{Emery:87,Monthoux:91}.
%Steadily for a period of 36 years, 100 -- 200~thousand experimental papers studying HTS were published.
Whence, a natural problem arises: 
\textit{which of these experiments can be considered crucial for our
understanding of the mechanism of HTS? What does the critical
temperature, say $T_c$, depend on and with what does it correlate?}

The answers to these questions do not seem to be within reach now than they were at the time of the HTS discovery. The first proposals~\cite{Anderson:87,Varma:87} led to numerous 
models~\cite{Emery:87,Zhang:88,Monthoux:91,Anderson_Th,Lee:06,Kivelson:15,Spalek:22,Arovas:22},
but no clear winner is in sight yet.
So far, it seems strange that the crucial experiment
to solve this problem is rather a numerical one,
a method which has never led to a Nobel prize in physics.
Almost 40 years after their development, High-$T_c$ cuprates
are still in the limelight of condensed matter by virtue of their
great potential application in many fields.
For every compound, $T_c$ depends on doping or the chemical potential.
Let us recall the well-known parabolic
approximation obtained by fitting
the experimental data of many high-$T_c$ cuprate materials~\cite{Tallon:91,Obertelli:92}
\be
 T_c/T_{c,\,\mathrm{max}}=1-82.6\, (\tilde p-0.16)^2
 \label{doping_dependence}
\ee
close to the optimal doping $\tilde p_\mathrm{opt}=0.16$ holes per Cu ion in the CuO$_2$ plane.
This maximum is far from the metal-insulator transition and close to this maximum 
high-$T_c$ cuprates are in a first approximation normal metals,
for which the electron band theory holds.
The \textit{ab initio} calculated Fermi surface is in excellent agreement with Angle Resolved Photo Emission Spectroscopy (ARPES) data~\cite{MishonovPenev:11}.
Moreover, even the BCS spectrum of the normal excited states is
observable in the case of Bi$_2$Sr$_2$CaCu$_2$O$_{8+\delta}$
\cite{MishonovPenev:11,MishIndPen:03}.
Additionally, close to the optimal doping
the pseudo gap, if any, is small and has a weak influence on $T_c$ and the thermodynamic properties of cuprates.

The nature of this doping dependence on the critical temperature 
\Eqref{doping_dependence} is not yet clear.
What causes $T_c$ to fall so swiftly with overdoping? 
It has been suggested that the pairing interaction must decrease, since
the electronic specific heat, \textit{i.e.} the electron density of
states, remains constant on the overdoped side. See, for example,
the paper by Storey at al.~\cite{Storey:07}
who consider the influence of the
saddle-point Van Hove singularity 
on the phase diagram of high-$T_c$ cuprates.
In the present paper we concentrate only on the optimal doping
studied by Pavarini et al~\cite{Pavarini:01}
assuming that
the dominant electron pairing mechanism is common for all cuprates.
Obviously, far from the optimal doping a lot of new physics becomes
important.

All these considerations allow us to use a classical approach of
electron band calculations and BCS pairing to identify the critical temperature 
$T_{c,\,\mathrm{max}}$ at optimal doping.
A preliminary qualitative analysis was performed 
a long time ago~\cite{BJP:11}. Here, we give a far more detailed and quantitative one.
The technical details of the 
Linear Combination of Atomic Orbitals (LCAO) approximation
of the electron band structure of the CuO$_2$ plane and the matrix elements of the Kondo interaction
in this approximation are given in Ref.~\cite{MishonovPenev:11}.

\section{Results}

All HTS cuprates contain stacks of CuO$_2$ planes, but how can the critical temperature $T_{c,\,\mathrm{max}}$
be so different even among optimally doped superconductors with $\tilde p=0.16$ holes per Cu ion~\cite{Tallon:91}?
A hint to the nature of HTS in cuprates in the
LCAO calculations by Andersen \textit{et
al.}~\cite{Andersen:95,Andersen:96} was advanced
by R\"{o}hler~\cite{Roehler:00,Roehler:00a},
who suggested that the hybridization between Cu$3d$ and Cu$4s$
orbitals is a crucial parameter to the CuO$_2$ plane.
This assertion was strongly confirmed by the remarkable correlation
between the Cu$4s$ energy level and the critical temperature
$T_{c,\,\mathrm{max}}$ from band calculations~\cite{Pavarini:01}.
It turns out that $T_{c,\,\mathrm{max}}$ for optimally doped cuprates
is tightly related to the dimensionless ratio $t^\prime/t$, where the
parameters $t$ and $t’$ determine the shape of the Fermi contour (FC)
via the very well-known and
widely used formula~\cite{Pavarini:01,MishonovPenev:11}
\begin{align}
&-2t\,[\cos p_x+\cos p_y]+4t^\prime\cos p_x \cos p_y =\mathrm{const.},
\label{t^prime_t}\\
&\; \, p_x\equiv a_0P_x/\hbar,\quad p_y\equiv a_0P_y/\hbar, 
\quad 
0 \le p_x,\, p_y \le 2\pi, \nn
\end{align}
where $a_0$ is the Cu-Cu interdistance (i.e. the lattice constant), 
and $\mathbf{P}$ is the electron quasi-momentum of motion in the CuO$_2$ plane. 

The correlation $T_{c,\,\mathrm{max}}$ versus
$t^\prime/t$~\cite{Pavarini:01} has gained a
broad recognition,
unfortunately as a curious empirical correlation 
without any microscopic theoretical understanding.
One of the goals of the present study is to present a microscopic
explanation of the correlation between 
$T_{c,\,\mathrm{max}}$ and the shape of the Fermi contour,
that is the mapping of the Fermi surface on a 2D plane.
The identification of $T_c$ is crucial to developing the theory 
of CuO$_2$ plane layered superconductors.
%Models and mechanisms of oxide superconductors 
%where important themes of many conferences.

%%%%%%%%%%%
\begin{figure}[h!]
\centering
\includegraphics[scale=0.5]{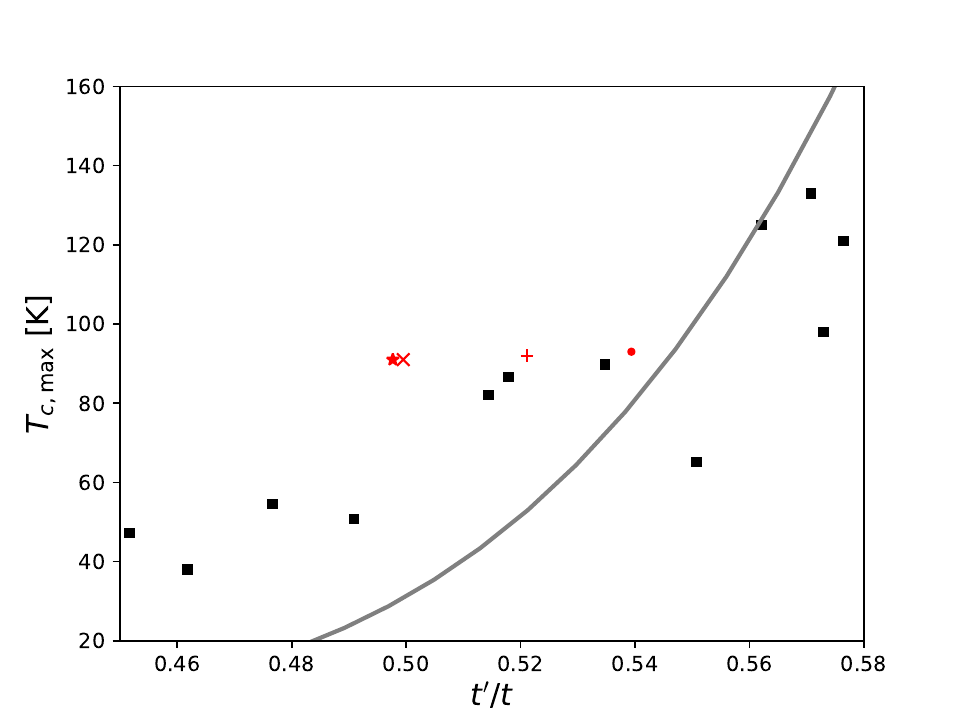}
\caption{
The correlation between the critical temperature
of optimally doped cuprates $T_{c,\,\mathrm{max}}$ ($_\blacksquare$)
and the dimensionless parameter  $t^\prime/t$ describing the shape of the Fermi contour is one of the most important properties of the physics 
of high-$T_c$ cuprates after Ref.~\cite{Pavarini:01} and our \textit{s-d} exchange calculation (solid line).
ARPES data is also included with
({\color{red}$\mathbf{\star}$}) by Zonno \textit{et al.} \cite[Fig.~5c]{Zonno:21}, 
({\color{red} $\mathbf{\times}$}) by Zhang \textit{et al.}~\cite[Fig.~1]{Zhang:14},
%for optimally doped Bi-2212, 
({\color{red} $\mathbf{+}$}) by Nakayama \textit{et al.}~\cite[Fig.~1]{Nakayama:07} and
({\color{red} $\mathbf{\bullet}$}) by Okawa \textit{et al.}~\cite[Fig.~1]{Okawa:10}.
%for optimally doped YBCO.
The pairing is created by an interaction
but the Fermi contour is a notion of 
independent electron band approximation.
What physics is hidden behind the
here reproduced correlation?
}
\label{Fig:rTc}
\end{figure}
%%%%%%%%

\begin{figure}[ht!]
\centering
\includegraphics[scale=0.5]{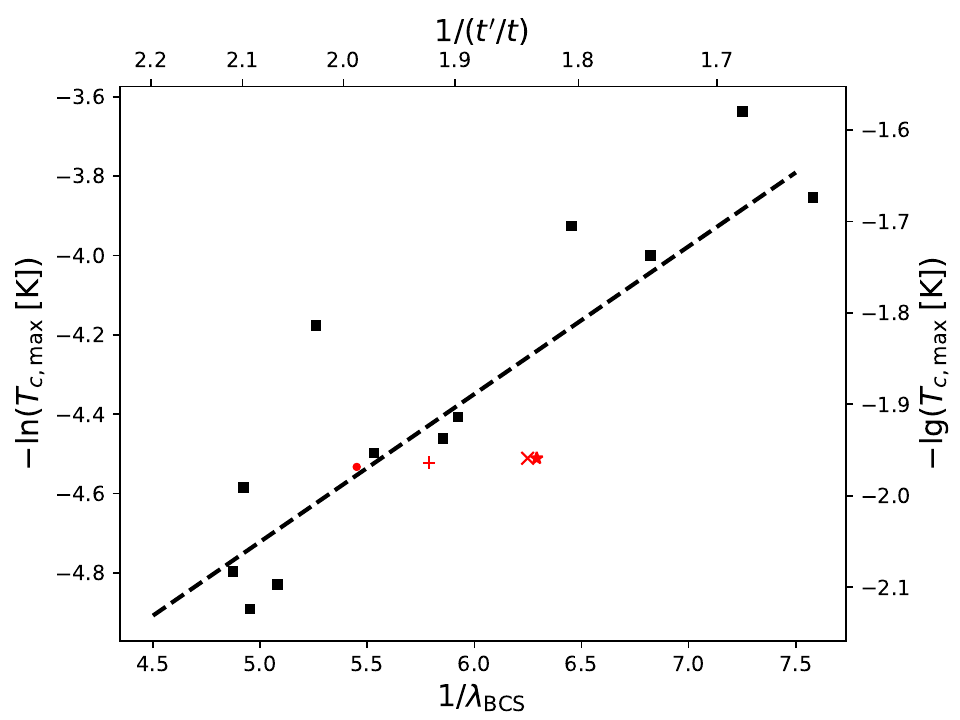}
\caption{
The linear dependence between
$\ln(T_{c,\,\mathrm{max}})$
and BCS coupling constant $\lambda$
for the Kondo interaction reveals that in and
acceptable accuracy pairing in CuO$_2$
plain can be described in the framework of
BCS scheme.
The high correlation coefficient of this linear regression gives a hint that the Kondo exchange amplitude $J_{sd}$ is similar for all cuprates.
Markers are the same as in \Fref{Fig:rTc}.
The continuous solid line is a theoretical curve that
serves as a guide to the eye.}
\label{Fig:lmbTc_}
\end{figure}

In \Fref{Fig:rTc}, we reproduce this correlation, 
as we add to the electron band calculations
new data for $t^\prime/t$ obtained by ARPES experiments,
a technique widely used for studying  
cuprates~\cite{Damascelli:03,Damascelli:04,Yu:20,Shen:21}.
$T_{c,\,\mathrm{max}}$ is determined according to the
parabolic doping dependence \Eqref{doping_dependence}.
The ratio $t^\prime/t$ is itself exactly a linear function
of the BCS coupling parameter $\lambda$
supposing the Kondo \emph{s-d} exchange amplitude $J_{sd}$ is constant,
i.e. $J_{sd}=5.593$~eV.
According to our classical BCS interpretation (solid line) in
\Fref{Fig:rTc},
this band-structure trend describes the
$T_{c,\,\mathrm{max}}$-$\lambda$ correlation for $J_{sd}$
approximately equal for all cuprates. 
The numerical value of $J_{sd}$ is determined from the slope of a linear
regression of the experimental data,
see FIG. \ref{Fig:lmbTc_}, obtained for high-$T_c$ superconductors whose Fermi contour
is accessible by ARPES and its shape is in good agreement
with electronic band calculation. 
In this case the correlation between $T_{c,\mathrm{max}}$ and 
shape of FC is rather an experimental relation.
Fitting the experimental data to a single universal curve indicates that an approximate dependence is revealed.

There is no doubt that for underdoped cuprates the Fermi contour is
not well defined, but the correlations suggested in
Ref.~\cite{Pavarini:01} pertain to optimally doped cuprates for which the electron band theory is well applicable. 
Thus, the perfect agreement between electron band calculations and the Fermi
contour observed experimentally by ARPES is a rather
convincing outcome.

According to the original consideration by Mott,
every metal becomes an insulator if we gradually increase the lattice constant.  
In brief, optimally doped cuprates are at first approximations normal
metals whose \textit{ab initio} calculated dispersion of the conduction band is in acceptable agreement with the experimentally observed one.
Concerning the statistical properties for optimally doped superconductors,
the BCS-Bogolyubov spectrum of the electron excitations 
has also been experimentally confirmed~\cite{Campuzano:18}.
Whence, the BCS trial function approach leads to
an acceptable estimate of the 
thermodynamic and kinetic properties of optimally doped cuprates,
whose pseudogap is too small to
perturb and break the BCS picture.
As for the normal state properties, such as strong anisotropy of the normal state scattering rate and the linear temperature dependence of the in plane resistivity,
they also have a conventional explanation in the framework of the normal metal theory if we consider the Kondo interaction~\cite{Kondo} as a scattering mechanism of normal charge carriers in a layered metal;
see for example Refs.~\cite{hotspot2022,arXiv_Pokrovsky} and references therein.

%%%%%%%%
%%%%%%%%%

\begin{figure}[ht]
\centering
\includegraphics[scale=0.5]{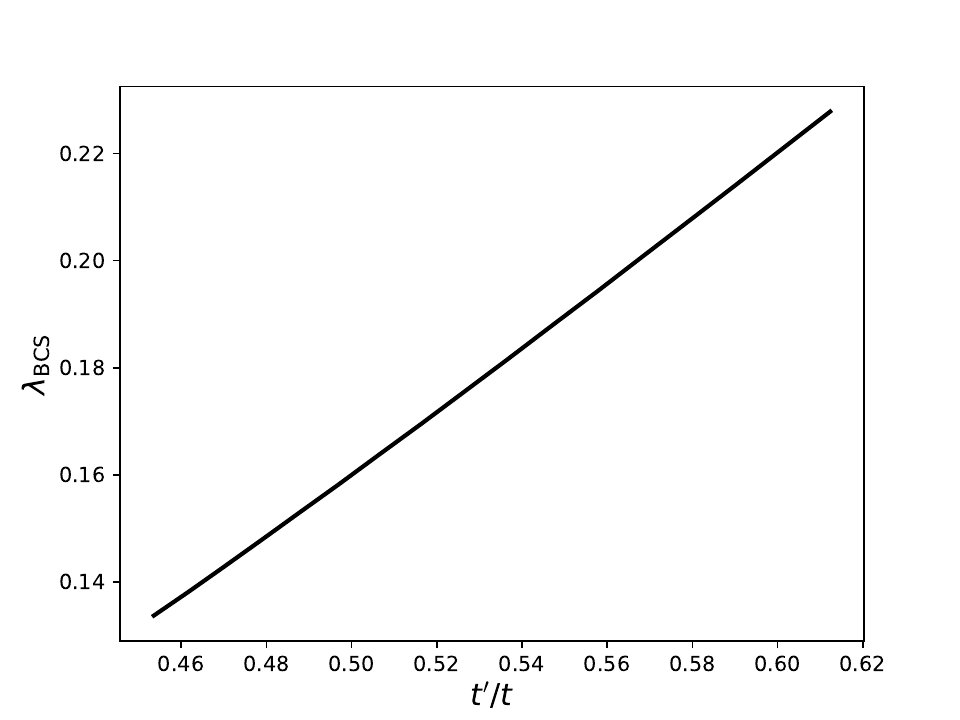}
\caption{The linear dependence between 
the coupling constant of the Kondo interaction
$\lambda_\mathrm{BCS}$ and the dimensionless
parameter $t^\prime/t$ determining the shape of Fermi contour at optimal hole doping is one of the main properties of the electronic structure of the CuO$_2$ plane.}
\label{Fig:tpt-lambda_}
\end{figure}

Often the choice of variables is crucial to further analysis. In
\Fref{Fig:lmbTc_}, the data are represented by a
logarithm of the ordinate and the reciprocal value of the abscissa,
i.e the correlation -$\ln T_c$ versus $t/t^\prime$ is depicted.
The straight line is 
obtained supposing that for different cuprates only the
position of  the 4$s$ level $\epsilon_s$ is different 
and all other parameters of the Hamiltonian are constants with $J_{sd}=5593$~meV.
Within this approximation, the coupling constant $\lambda$ 
is  a linear function of $t^\prime/t$ depicted in \Fref{Fig:tpt-lambda_}.
The slope of the linear regression (dashed line) of the data from Ref.~\cite{Pavarini:01} ($_\blacksquare$) together with the ARPES data included in \Fref{Fig:rTc} determines the value of $J_{sd}$.
%({\color{red}$\mathbf{\star}$}) by Zonno \textit{et al.} \cite[Fig.~5c]{Zonno:21} and
%({\color{red} $\mathbf{\times}$}) by Zhang \textit{et al.}~\cite[Fig.~1]{Zhang:14}
%for optimally doped Bi-2212, and
%({\color{red} $\mathbf{+}$}) by Nakayama \textit{et al.}~\cite[Fig.~1]{Nakayama:07} and
%({\color{red} $\mathbf{\bullet}$}) by Okawa \textit{et al.}~\cite[Fig.~1]{Okawa:10}
% for optimally doped YBCO.
It is worth mentioning that the straight line is obtained via linear
regression analysis of these data summing up decades of development
of the physics of HTS. Moreover, 
the high correlation coefficient $\rho\approx0.85$
rises the simple problem:
\textit{to find an approximation theoretically explaining the linear dependence 
between $\ln  T_{c,\,\mathrm{max}}$ and $t/t^\prime$.}
The high number of synthesized HTS cuprates demonstrates that we have to search for some simple mechanism reliably hidden in the textbooks on solid state physics.

%%%%%%%%%%%
%%%%%%%%

In the 21-st century, the consensus
that HTS is created by specific exchange processes has gradually started
to emerge (see e.g. Ref.
\cite{Emery:87,Zhang:88,Monthoux:91,Anderson_Th,Lee:06,Kivelson:15,Spalek:22,Arovas:22}
and references therein), whence the first candidate is 
the \textit{s-d} Kondo interaction, which describes many phenomena
related to exchange interaction inherent to the copper ion.
Obviously, the simplest possibility is to incorporate the Kondo interaction 
between itinerant electrons in the standard BCS scheme,
as for the band wave functions we use the LCAO approximation.
The system of notions and notations closely follows those introduced in the monograph
\cite{MishonovPenev:11}.
See also our recent paper on hot/cold spot phenomenology along the
Fermi contour in cuprates~\cite{hotspot2022} and the possible zero sound
in layered perovskites~\cite{sound2022}.
A similar study on the correlation between
the pairing interaction and electron-electron scattering in the normal phase
was carried out by Hussey et al.~\cite{Hussey:11}.
Similar relations are well-known in the phenomena
dominated by exchange interaction, such as Heavy Fermions \cite{Hewson:93}.

\section{Discussion}

The sensitivity of $T_{c,\,\mathrm{max}}$ with respect to the Cu4$s$ level reveals that $\epsilon_s$ (the energy
position of  the 4$s$ level) is so important, since it determines
the  $s$-$d$ hybridization which is the main detail of the Kondo
\textit{s-d} exchange interaction yielding the BCS coupling constant 
$\lambda=V_0 \rho_\mathrm{_F}$.
The pairing mechanism is revealed with the help of the remarkable
correlation between $T_{c,\mathrm{max}}$ and $t^\prime/t$~\cite{Pavarini:01} which has not
attracted the deserving theoretical interest so far.

Here, we report how the \textit{s-d} Kondo interaction
incorporated in the BCS theory describes the well-known correlation between the critical temperature and the shape of the Fermi contour. 
This correlation describes the difference in the critical temperature of many optimally doped cuprates, while
%and due to lack of the alternative explanations in the past 20 years gives a hint that this will remain true in the next 20 years;
the \textit{s-d} Kondo interaction has always been well-known to
theorists studying exchange magnetism and kinetics of various processes in condensed matter. Besides, the \textit{s-d} interaction
describes many of the properties of the normal phase of layered
cuprates, linear temperature dependence of the resistivity and cold
spots along the nodal directions, to name a few.
This provides a reliable basis for further studies of exchange processes related to the Cu ion in the CuO$_2$ plane.
Midst the unresolved problems in this direction we wish to mention $2\Delta_\mathrm{max}/T_c$
versus $T_c$; the \textit{s-d} interaction gives much weaker
dependence even with an opposite sign \cite{Wei:98,Handbook}.

\section{Methods. Kondo $s$-$d$ exchange interaction and LCAO method incorporated in the BCS theory}

\subsection{BCS gap equation for \textit{s-d} pairing mechanism}
The theory of $\textit{s-d}$ exchange mediated superconductivity
was developed many years ago in the monograph~\cite{MishonovPenev:11}.
This pairing mechanism naturally explains the gap anisotropy,
temperature dependence of the gap, heat capacity and penetration depth.
But the main problem is whether this pairing mechanism can explain 
in the same natural manner the correlations between
$T_{c,\mathrm{max}}$ and dimensionless $t^\prime/t$
describing the shape of the Fermi contour at fixed doping level $\tilde p$.

The LCAO approximation of the constant energy curves (CEC) and the 
Fermi contour \Eqref{spectrum_} are straightforwardly obtained from
the determinant of the LCAO equation for the band structure
of the 4-band model;
Eq. \eqref{spectrum_} provides the microscopic derivation of
the expression for
CEC, \Eqref{t^prime_t}, postulated as a property of a quadratic lattice
ignoring atomic wave functions.
The analysis of the influence of all parameters of the LCAO model
(single site energies and transfer amplitudes) revealed that the shape
of the Fermi contour parameterized by the dimensionless parameter
used in the paper by Pavarini et al~\cite{Pavarini:01}
$\mathcal{F}\equiv t^\prime/t$
is most sensitive to the position of the Cu4s state $\epsilon_s$.
This sensitivity of the shape of the Fermi surface and the correlation with 
$T_{c,\mathrm{max}}$ suggested that the Cu$4s$ state is crucial to
the pairing in the CuO$_2$ plane.
Having in mind that
copper-oxide high-$T_c$ superconductivity is induced by an appropriate superexchange
mechanism, it was naturally to identify the exchange amplitudes 
related to the Cu$4s$ state.
Actually, there is only one possible parameter, namely $J_{sd}$ introduced as a phenomenology
by Zener even before BCS the theory was developed.
In addition, phenomenologically analyzing a completely different
problem, Kondo concluded the conclusion that $J_{sd}$ can describe
antiferromagnets as well.
Furthermore, when this interaction is incorporated in the BCS theory,
it will promote pairing in 
the singlet channel similarly to phonon-electron superconductors.

Now we outline the equations 
describing the link between the experimental results and our calculations.
First, let us recall the BCS equation for anisotropic gap superconductors~\cite{MishonovPenev:11}
\begin{align}
\label{BCS_gap_equation_}
& 2J_{sd}\,
\overline{\frac{\chi_\mathbf{p}^2}{2E_\mathbf{p}}
\tanh\left(\frac{E_\mathbf{p}}{2T}\right)} =1,
\qquad E_\mathbf{p}=\sqrt{\eta_\mathbf{p}^2+\Delta_\mathbf{p}^2},\\
&\eta_\mathbf{p}=\epsilon_\mathbf{p}-\eF,
\qquad \Delta_\mathbf{p}=\Xi(T)\,\chi_\mathbf{p},
\qquad \mathbf{p}=a_0 \mathbf{P}/\hbar,
\nn\\
&\overline{F(\mathbf{p})} \equiv \int\limits_0^{2\pi}\int\limits_0^{2\pi}
F(p_x,p_y)\,\frac{\md p_x \md p_y}{(2\pi)^2},
\qquad p_x,\,p_y\in (0,\, 2\pi),
\nn
\end{align}
where, the superconducting gap $\Delta_\mathbf{p}$
is factorized to a product of a temperature dependent order parameter 
$\Xi(T)$ and the momentum dependent gap anisotropy function 
$\chi_\mathbf{p}$,
$\epsilon_\mathbf{p}$ is the energy dispersion of the conduction band,
$\eF$ is the Fermi energy, and the overline denotes momentum integration in the Brillouin zone.
The gap anisotropy function is determined by the electron interaction
Hamiltonian.

Here, we emphasize that the general BCS theory for anisotropic gap superconductors was originally developed by Pokrovsky~\cite{Pokr:61} within the weak coupling limit.
In this case the anisotropy function $\chi_\mathbf{p}$ is the eigenfunction of the pairing interaction.
However, for the Kondo interaction in the CuO$_2$ plane the pairing
attraction is naturally factorizable
$\propto\chi_\mathbf{p}\chi_\mathbf{q}$ and the Pokrovsky results are
a consequence of the BCS scheme applied as a trial function and in
this case, they have a broader area of applicability than the weak coupling approximation.

\subsection{LCAO approximation and $s$-$d$ exchange interaction}

Pairing by \emph{s-d} interaction is an old idea,
see, for example Ref.~\cite{Tonchev:80,Bogolyubov:1981,chamati1992a,chamati1994b} and references therein.
LCAO method applied to cuprate structures
gives the unique possibility to obtain the analytical expression for the gap anisotropy function, the details in the derivation are given in \cite[Eq.~(2.31)]{MishonovPenev:11}
%\begin{align}
%\chi_\mathbf{p} =\,&
%4\varepsilon_\mathrm{p}t_{sp}t_{pd}(x-y)
%\nn\\
%&\times\left[
%\varepsilon_s\varepsilon_\mathrm{p}^2-4\varepsilon_\mathrm{p}t_{sp}^2\,(x+y)
%+32t_{pp}\tau_{sp}^2\,xy
%\right]\nn\\
%&\times\left\{
%\left[4\varepsilon_\mathrm{p}t_{sp}t_{pd}\,(x-y)\right]^2\right.\nn\\
%&\;\qquad
%+\left[\varepsilon_s\varepsilon_\mathrm{p}^2-4\varepsilon_\mathrm{p}t_{sp}^2\,(x+y)
%+32t_{pp}\tau_{sp}^2\,xy \right]^2\nn\\
%&\;\qquad
%+4x\left[(\varepsilon_s\varepsilon_\mathrm{p}-8\tau_{sp}^2y)t_{pd}\right]^2\nn\\
%&\;\qquad
%\left.+4y\left[(\varepsilon_s\varepsilon_\mathrm{p}-8\tau_{sp}^2x)t_{pd}\right]^2
%\right\}^{-1},
%\label{chi_analytical_}
%\end{align}
\begin{widetext}
\be
\chi_\mathbf{p} =
\dfrac{4 \varepsilon_\mathrm{p}t_{sp}t_{pd}(x - y)
\left[
\varepsilon_s\varepsilon_\mathrm{p}^2-4\varepsilon_\mathrm{p}t_{sp}^2 (x  +  y)
+32t_{pp}\tau_{sp}^2 xy
\right]}
{
\left[4\varepsilon_\mathrm{p}t_{sp}t_{pd} (x \! - \! y)\right]^2 \!
+\left[\varepsilon_s\varepsilon_\mathrm{p}^2-4\varepsilon_\mathrm{p}t_{sp}^2 (x \! + \! y)
+ 32t_{pp}\tau_{sp}^2 xy \right]^2\!\!
+4x\left[(\varepsilon_s\varepsilon_\mathrm{p} \! - \! 8\tau_{sp}^2y)t_{pd}\right]^2 \!\!
+4y\left[(\varepsilon_s\varepsilon_\mathrm{p} \!- \! 8\tau_{sp}^2x)t_{pd}\right]^2\!\!
} \,\, ,
\label{chi_analytical_}
\ee
\end{widetext}
where the multiplier
$$(x-y)=-\frac12\left[\cos(p_x)-\cos(p_y)\right]\propto
p_x^2-p_y^2,
$$
when $p_x^2+p_y^2\ll1$,
describes the $d$-type symmetry of the superconducting gap and
\begin{align}
& \varepsilon_s=\epsilon-\epsilon_s,\qquad
\varepsilon_d=\epsilon-\epsilon_d,\qquad
\varepsilon_\mathrm{p}=\epsilon-\epsilon_\mathrm{p},\nn \\
& \tau_{sp}^2=t_{sp}^2-\frac12\varepsilon_st_{pp},\quad
x=\sin^2\left(\frac12p_x\right),\quad y=\sin^2\left(\frac12p_y\right).\nn
\end{align}
Here $\epsilon$ is the electron energy,
$\epsilon_d$ is the energy of Cu3$d_{x^2-y^2}$ level, 
$\epsilon_s$ is the energy Cu4$s$, and 
$\epsilon_p$ is the energy of oxygen O2p$_x$ and O2p$_y$ levels.
The subscripts of the transfer integrals $t_{sp}$, $t_{pd}$ and $t_{pp}$
indicate the interacting neighboring atomic orbitals.
The momentum dependent hybridization function $\chi_\mathbf{p}$
stands for the probability for one
electron from the conduction band to be simultaneously in Cu4$s$
and Cu3$d_{x^2-y^2}$.
While the LCAO Hamiltonian is described in many textbooks,
the Hamiltonian of the Kondo interaction deserves a more detailed description.
The exchange amplitude $J_{sd}$ explains the correlated hopping localized around the single Cu ion.
One 3$d$ electron jumps to the 4$s$ orbital ,
while a 4$s$ electron reaches the 3$d$ orbital.
Let $\hat{S}_{\mathrm{n},\alpha}$ be the annihilation operator for one electron
with spin projection $\alpha$ in the Cu4$s$ state in the $\mathbf{n}=(n_x,\,n_y)$
elementary cell, where $n_x,\,n_y=0,\,\pm1,\,\pm2,\dots\,$, and
$\hat D_{\mathrm{n},\beta}^\dagger$ is the Fermi creation operator for an electron
in the Cu3$d_{x^2-y^2}$ state with spin projection $\beta$.
For the surrounding environment there is no change of the electrostatic correlation. This two-electron process is a consequence of the correlated hopping in which the electrostatic repulsion is minimized.
When correlations are so strong, they are simply included in the effective Hamiltonian.
If we write this in the second quantization representation, we write a 4-fermion term
with 2 creation and 2 annihilation operators for every Cu ion and
additionally we have to sum over all transition metal ions in the crystal.
Whence, way the Kondo \textit{s-d} exchange Hamiltonian considered
here as a pairing interaction reads
\be
\hat H_\mathrm{Kondo}=-J_{sd}\sum_{\mathbf{n},\,\alpha, \,\beta}
\hat{S}_{\mathbf{n},\,\alpha}^\dagger\hat{D}_{\mathbf{n},\,\beta}^\dagger
\hat{S}_{\mathbf{n},\,\beta}
\hat{D}_{\mathbf{n},\,\alpha}.
\label{Eq.Kondo}
\ee
For more extended consideration of this and other exchange Hamiltonians 
see Ref.~\cite[Eqs.~(2.9)]{MishonovPenev:11}
and for the BCS reduced Hamiltonian see Ref.~\cite[(2.26)]{MishonovPenev:11}.
In the microscopic consideration in the physics of magnetism of
transition metal ions, the calculation of the antiferromagnetic Kondo
amplitude $J_{sd}$ is tightly related to the strong Coulomb interaction $U_{dd}$ of two electrons in the $3d$ orbital.
Hence, the phenomenological Kondo exchange takes into account strong
electron correlations for specific purposes.
In this sense, high-$T_c$ cuprates are representatives of strongly correlated quantum matter.
The antiferromagnetic exchange amplitude $J_{sd}$
arises from the interplay of intermediate processes
involving O2$p$ orbitals and the Hubbard repulsion.
This is not an intra-atomic process but a property
of the Cu ion and oxygen ligands.
The cluster of Cu ions and the 4 adjacent oxygen ions is common to all
CuO$_2$ plane superconductors, and one can expect 
close numerical values for the $J_{sd}$ exchange amplitude in all
high-$T_s$ cuprates.
The intra-atomic Coulomb repulsion between the Cu4$s$ and Cu3$d$
electrons is related to the Hund rule for the maximal spin in single atomic spectra.

Let us explain qualitatively how the Kondo interaction 
leads to superconductivity.
In general, the \textit{s-d} Zener Hamiltonian describes the effective
two electron interaction when two electrons are simultaneously
in one and the same transition ion in 3$d$ and 4$s$ states.
The sign of the exchange amplitude $J_{sd}$ describes
a decrease of the energy when the electron spins are opposite,
corresponding to an antiferromagnetic interaction of the effective
two electrons. In brief, when Kondo the interaction is incorporated
in the BCS scheme it gives singlet Cooper pairs.

The Kondo Hamiltonian
\Eqref{Eq.Kondo} describes the two-electron exchange between
the Cu4$s$ and Cu3$d_{x^2-y^2}$ states in one and the same 
Cu ion. 
In a way, our theory is similar to the \textit{t-J} model, yet the
LCAO-Kondo Hamiltonian is adequate for the treatment 
of electron processes in the CuO$_2$ plane. 
Let's point out, the notation $t$ comes from the notion
\textit{transfer amplitude}.
Our $t_{ps}$, $t_{pd}$ and $t_{pp}$ are
transfer amplitudes between real atomic wave functions.

Introducing the electron $\hat X_{\mathbf{n},\alpha}$ 
annihilation operator for the O$2p_x$ orbital at $\mathbf{n}= (n_x,\,n_y)$ 
elementary cell with spin projection $\alpha$, and analogously 
the corresponding electron $\hat Y_{\mathbf{n},\alpha}^\dagger$
creation operator, the LCAO Hamiltonian of the CuO$_2$ plane~\cite[Fig.~1.1, Eq.~(1.2)]{MishonovPenev:11} reads
%\begin{align}
%\hat H_\mathrm{LCAO} \! = \! \sum_{\mathbf{n},\alpha} &
%\left\{D_{\mathbf{n},\alpha}^\dagger
%\left[-t_{pd}( -\hat X_{\mathbf{n},\alpha}
%+\hat X_{x-1,y,\alpha}
%\right.\right.  \\
%& 
%\left. \qquad\qquad\qquad
%+\hat Y_{\mathbf{n},\alpha}
%-\hat Y_{x,y-1,\alpha})+\epsilon_d\hat D_{\mathbf{n},\alpha} \right] \nn\\
%& 
%+\hat S_{\mathbf{n},\alpha}^\dagger
%[-t_{sp} (-\hat X_{\mathbf{n},\alpha}+\hat X_{x-1,y,\alpha} \nn \\
%& \qquad\qquad\quad \;\;
%-\hat Y_{\mathbf{n},\alpha}+\hat Y_{x,y-1,\alpha})
%+\epsilon_s\hat S_{\mathbf{n},\alpha}]
%\nn \\
%& 
%+\hat X_{\mathbf{n},\alpha}^\dagger
%[-t_{pp} (\hat Y_{\mathbf{n},\alpha}-\hat Y_{x+1,y,\alpha} \nn \\
%&  \qquad\qquad\quad 
%-\hat Y_{x,y-1,\alpha}+\hat Y_{x+1,y-1,\alpha}) \nn \\
%&  \qquad\quad 
%-t_{sp} (-\hat S_{\mathbf{n},\alpha}+\hat S_{x+1,y,\alpha}) \nn \\
%& \qquad\quad 
%-t_{pd} (-\hat D_{\mathbf{n},\alpha}+\hat D_{x+1,y,\alpha})
%+\epsilon_p \hat X_{\mathbf{n},\alpha}] \nn \\
%&
%+\hat Y_{\mathbf{n},\alpha}^\dagger
%[-t_{pp} (\;\; \hat X_{\mathbf{n},\alpha}-\hat X_{x-1,y,\alpha} \nn \\
%&  \qquad\qquad\quad \;
%-\hat X_{x,y+1,\alpha}+\hat X_{x-1,y+1,\alpha}) \nn \\
%&  \qquad\quad 
%-t_{sp} (-\hat S_{\mathbf{n},\alpha}+\hat S_{x,y+1,\alpha}) \nn \\
%& \qquad\quad 
%-t_{pd} (\;\; \hat D_{\mathbf{n},\alpha}-\hat D_{x,y+1,\alpha})
%+\epsilon_p \hat Y_{\mathbf{n},\alpha}]
%\Bigl. \Bigr\}.
%\nn
%\end{align}
\begin{widetext}
\begin{align}
\hat H_\mathrm{LCAO} & \! =  \! \sum_{\mathbf{n},\alpha} 
\left\{D_{\mathbf{n},\alpha}^\dagger
\left[-t_{pd}( -\hat X_{\mathbf{n},\alpha}
+\hat X_{x-1,y,\alpha}
+\hat Y_{\mathbf{n},\alpha}
-\hat Y_{x,y-1,\alpha})+\epsilon_d\hat D_{\mathbf{n},\alpha} \right] 
\right.
\nn
\\
& 
+\hat S_{\mathbf{n},\alpha}^\dagger
[-t_{sp} (-\hat X_{\mathbf{n},\alpha}+\hat X_{x-1,y,\alpha} 
-\hat Y_{\mathbf{n},\alpha}+\hat Y_{x,y-1,\alpha})
+\epsilon_s\hat S_{\mathbf{n},\alpha}]
\nn \\
& 
+\hat X_{\mathbf{n},\alpha}^\dagger
[-t_{pp} (\hat Y_{\mathbf{n},\alpha}\! - \! \hat Y_{x+1,y,\alpha} 
\! -\! \hat Y_{x,y-1,\alpha} \! + \! \hat Y_{x+1,y-1,\alpha}) 
\! - \! t_{sp} (-\hat S_{\mathbf{n},\alpha} \! +\! \hat S_{x+1,y,\alpha}) 
\! -\! t_{pd} (-\hat D_{\mathbf{n},\alpha} \! +\! \hat D_{x+1,y,\alpha})
\! +\! \epsilon_p \hat X_{\mathbf{n},\alpha}] \nn \\
&
+\hat Y_{\mathbf{n},\alpha}^\dagger
[-t_{pp} (\hat X_{\mathbf{n},\alpha} \! -\! \hat X_{x-1,y,\alpha} 
\! - \! \hat X_{x,y+1,\alpha} \! +\! \hat X_{x-1,y+1,\alpha}) 
\! - \! t_{sp} (-\hat S_{\mathbf{n},\alpha} \!+ \! \hat S_{x,y+1,\alpha})
\! - \! t_{pd} ( \hat D_{\mathbf{n},\alpha} \! - \! \hat D_{x,y+1,\alpha})
\! +\! \epsilon_p \hat Y_{\mathbf{n},\alpha}]
\Bigl. \Bigr\}.
\label{4_band_LCAO}
\end{align}
\end{widetext}

Optimally doped cuprates are metals far from the metal-insulator transition
and thus electron band calculations
may be used with admissible accuracy.
Moreover, the relevant to the superconductivity bands can be approximated 
very well by the tight-binding method.
Roughly speaking, in this approximation we have the Hilbert space spanned on
the Cu4$s$,  Cu3$d_{x^2-y^2}$, O2$p_x$, and O2$p_y$ atomic states
in the CuO$_2$ plane.
For details of a pedagogical consideration of the LCAO Hamiltonian 
of CuO$_2$ plane see 
Ref.~\cite[Sec.~1.3, Eqs.~(1.1-1.12)]{MishonovPenev:11}.
For CEC, which is the Fermi contour
for $\epsilon=\eF$, the simple analytical equation
\be
\mathcal{A}xy+\mathcal{B}(x+y)+\mathcal{C}=0
\label{spectrum_}
\ee
is derived, from where, we have three energy dependent functions
\begin{subequations}
\label{ABC_}
\begin{align}
\mathcal{A}(\epsilon) &= 16(4t_{pd}^2t_{sp}^2+2t_{sp}^2t_{pp}\varepsilon_d
-2t_{pd}^2t_{pp}\varepsilon_s
-t_{pp}^2\varepsilon_d\varepsilon_s),
\\
\mathcal{B}(\epsilon) &= -4\varepsilon_\mathrm{p}(t_{sp}^2\varepsilon_d+t_{pd}^2\varepsilon_s),
\\
\mathcal{C}(\epsilon) &= \varepsilon_d\varepsilon_s\varepsilon_\mathrm{p}^2
\end{align}
\end{subequations}
with their respective energy derivatives
\begin{subequations}
\begin{align}
\mathcal{A}^\prime(\epsilon)  &= 16\,\left[2(t_{sp}^2-t_{pd}^2)
-(\varepsilon_d+\varepsilon_s)t_{pp}\right]t_{pp}, \\
\mathcal{B}^\prime(\epsilon)  &= -4(t_{sp}^2\varepsilon_d+t_{pd}^2\varepsilon_s)
-4(t_{sp}^2+t_{pd}^2)\,\varepsilon_p,\\
\mathcal{C}^\prime(\epsilon)  &=\left[\left(\varepsilon_s+\varepsilon_d\right)\varepsilon_p
+2\varepsilon_s\varepsilon_d\right]\varepsilon_p.
\end{align}
\end{subequations}
\Eqref{spectrum_} allows us to express CEC explicitly
\be
p_y=2\arcsin\sqrt{y},\qquad
0\le y=-\frac{\mathcal{B}x+\mathcal{C}}{\mathcal{A}x+\mathcal{B}}
\le1,
\ee
making possible the calculation of the band structure by numerical Newton method.
Finally, we have the convenient expression for the Fermi contour averaging
denoted by brackets $\langle\cdots\rangle$ as a linear integral from $p_x$
\begin{subequations}
\begin{align}
&
\langle f(\mathbf{p})\rangle
=\dfrac{\displaystyle\oint f(\mathbf{p})\dfrac{\md p_l}{v_\mathrm{_F}}}
{\displaystyle\oint\dfrac{\md p_l}{v_\mathrm{_F}}},
\quad v
\equiv\left|\frac{\partial \epsilon_\mathbf{p}}{\partial \mathbf p}\right|,\quad
V=\frac{a_0}{\hbar}v, 
\label{velocity_}
\\&
v(\mathbf{p})=\frac{\sqrt{(\mathcal{A}y+\mathcal{B})^2(1-x)x
                                     +(\mathcal{A}x+\mathcal{B})^2(1-y)y}}
{\mathcal{A}^\prime xy+\mathcal{B}^\prime(x+y)+\mathcal{C}^\prime},
\nn\\&
\md p_l=\sqrt{1+
\frac{(1-x)x}{(1-y)y}\left(\frac{\mathcal{A}y+\mathcal{B}}{\mathcal{A}x+\mathcal{B}}\right)^{2}} \; \md p_x, \nn\\
&
\rho_\mathrm{_F}=\frac{1}{(2\pi)^2}\oint\frac{\md p_l}{v_\mathrm{_F}}
=\overline{\delta(\epsilon_\mathbf{p}-\eF)},
\end{align}
\end{subequations}
or
\begin{subequations}
\be
\langle f(\mathbf{p})\rangle=\overline{f(\mathbf{p})\delta(\epsilon_\mathbf{p}-\eF)}
/\overline{\delta(\epsilon_\mathbf{p}-\eF)}.
\ee
\end{subequations}
%The brackets $\langle \dots\rangle$ describe averaging on the Fermi surface.
The derivative with respect the phases $v$ has dimension of energy,
and the velocity in usual units is denoted by $V$. The density of states 
$\rho_\mathrm{_F}$
per spin, unit cell 
and Cu atom at the Fermi level has dimension of inverse energy.
These algebraic results provide practical for numerical
analysis expressions for all variables of the BCS theory.
In the next subsection we provide only the results for the critical temperature.

\subsection{Main results of BCS scheme applied to anisotropic gap superconductors}

By averaging  the square of the gap anisotropy function along the Fermi contour
$\langle \chi^2\rangle$, we calculate the pairing energy $V_0$
and the dimensionless BCS coupling constant $\lambda$.
Additionally, the appropriately introduced $\chi_\mathrm{av}$
and rescaled gap anisotropy $\tilde\chi_\mathbf{p}$
alleviates the analysis of $2\Delta_\mathrm{max}/T_c$ ratio.
The analysis of the BCS anisotropic gap \Eqref{BCS_gap_equation_} gives
\begin{subequations}
\begin{align}
&T_c=\frac{2\gamma}{\pi}\,E_\mathrm{C}\, \exp(-1/\lambda), 
\quad
-\ln \frac{T_c}{E_\mathrm{C}}\!=\!\lambda^{-1}\!+\!\mathrm{const},
\label{TcBCS_}\\
& \lambda\equiv V_0\rho_\mathrm{_F}=2J_{sd}\langle \chi^2\rangle\rho_\mathrm{_F}, \quad
V_0 \equiv 2 J_{sd}  \langle \chi^2\rangle,
\label{lambdaBCS_}\\
&\tilde \Xi(0)=2E_\mathrm{C}\,\exp(-1/\lambda),\quad
\frac{2\tilde\Xi(0)}{T_c}=\frac{2\pi}{\gamma}\approx 3.53,
\label{pi_gamma}
\\
&\Delta_\mathbf{p}(T)=\tilde\Xi(T)\tilde\chi_\mathbf{p}=\Xi(T)\chi_\mathbf{p},
\quad \tilde \chi_\mathbf{p}=\frac{\chi_\mathbf{p}}{\chi_\mathrm{av}}\\&
\chi_\mathrm{av}
\equiv\exp\left\{\frac{\langle\chi_\mathbf{p}^2\ln|\chi_\mathbf{p}|\rangle}{\langle\chi_\mathbf{p}^2\rangle}\right\},
\quad
\frac{\Delta_\mathbf{p}(T)}{\Delta_\mathbf{p}(0)}
=\frac{\Xi(T)}{\Xi(0)}.
\end{align}
\end{subequations}
With the so defined dimensionless coupling
constant for the Kondo interaction $\lambda$,
we have re-derived the well-known BCS relation between 
the coupling constant and critical temperature. 
For the sake of comparison, we reproduce the well-known BCS relation for
isotropic superconductors equation \cite[Eq.~(40.4)]{LL9}
and \cite[Eq.~(34.23)]{AbrGorDzya}
\be
T_c=\gamma\Delta_0/\pi=0.5669329\,\Delta_0.
%{\magenta\ln\gamma= \mathrm C}.
\ee
The energy parameter $2 J_{sd}  \langle \chi^2\rangle$
is analogous to $V_0$ in the BCS theory.
The retardation effects are negligible for 
$T_c/E_\mathrm{C}\ll1$, i.e. when $\exp(-1/\lambda)\ll1$.
When $T_c$ is high enough the pseudo-potential of the Coulomb repulsion $\mu^*$
as other pairing mechanisms with different symmetries gives 
only perturbative corrections to $T_c$ and can be neglected at first approximation.
To calculate the critical temperature using \Eqref{TcBCS_}
and to represent the result in the standard BCS form,
we calculate the energy parameter 
\begin{align}
&\ln E_\mathrm{C}=\lim_{\epsilon\rightarrow 0}\left [\ln \epsilon
+\overline{\theta(|\eta_\mathbf{p}|>\epsilon)\chi_\mathbf{p}^2/|\eta_\mathbf{p}|}
/(2\langle\chi^2\rangle\rho_\mathrm{_F})
\right ],\nn\\
&E_\mathrm{C}\equiv\lim_{\epsilon\rightarrow 0}
\left [\epsilon\exp\left\{
\overline{\theta(|\eta_\mathbf{p}|>\epsilon)\chi_\mathbf{p}^2/|\eta_\mathbf{p}|}
/(2\langle\chi^2\rangle\rho_\mathrm{_F})
\right\} \right ].
\label{Euler-Asynchronism_energy_}
\end{align}
To this end, we use a numerical integration from the analytical
gap anisotropy $\chi_\mathbf{p}$ from 
\Eqref{chi_analytical_}.
This definition is similar to that of the
Euler-Mascheroni constant
\begin{align}
\mathrm{C}&\equiv\ln\gamma
=\lim_{N\rightarrow\infty}
\left(\ln N-\sum_{k=1}^N\frac1{k}\right)\\
&\approx 0.577215649\approx\ln (1.781052)
\end{align}
and one can call $E_\mathrm{C}$
Euler-Mascheroni energy.
We would like to point out that similar mathematics often appears in
quantum field theory.
%for example, $\zeta$-function of some operator $\hat A$
%with eigenvalues $a_n$ and eigenvectors $\vert n\rangle$
%\cite[Chap.~3, Sec.~5, Eq.~(5.2)]{Ramond:81}
%\begin{align}
%&
%\zeta_{\hat A}(s) =\sum_n \frac1{a_n^s}, \qquad
%\zeta_{\hat A}(s,z)=\sum_n \frac1{(a_n+z)^s},\\
%&
%\hat A \vert n\rangle= a_n\vert n\rangle, \qquad
%a_n>0, \nn
%\end{align}
%or even Hurwitz $\zeta(s,z)$-function of the same operator,
%cf. Ref.~\cite[Chap.~5, Eq.~(5.182)]{MishonovPenev:11}.

To analyze the parameter $E_\mathrm{C}$, as a simple example,
one can recall the classical BCS model
when the pairing interaction is isotropic 
and energy independent in a narrow energy band,
formally if $\chi=\theta(E_\mathrm{D}-|\eta|)$,
we have to calculate the integral
\be
E_\mathrm{D}
=\epsilon \exp\left(\int_\epsilon^{E_\mathrm{D}}
\frac{\mathrm{d}\eta}{\eta}
\right).
\ee
$E_\mathrm{C}$
can be considered as a generalization of the
Debye cutoff energy $E_\mathrm{D}$ of the
classical BCS isotropic phonon model.
When the masses of the atoms $M_a$ are slightly different
for phonon pairing,
the isotope effect reveals that the atomic masses affect the pre-exponent.
According to \Eqref{TcBCS_},
$T_c\propto E_\mathrm{D}\propto 1/\sqrt{M_a}$.
While for the $s$-$d$ exchange mechanism,
the Cu4$s$ level changes the coupling constant 
$-\ln T_c\propto 1/\lambda\sim \epsilon_s$.

The simplest illustration of the energy parameter $E_\mathrm{D}$
gives the isotropic gap $\chi_\mathbf{p}=1$
and the parabolic energy dispersion $\epsilon_\mathbf{p}=P^2/2m_e m_\mathrm{eff}$
in the two dimensional case when for charge carriers per plaquette and
fixed spin, we have
\be
N(P_\mathrm{\!_F})=\frac{\pi P_\mathrm{\!_F}^2}{(2\pi\hbar)^2}a_0^2,\qquad
\eF=\frac{P_\mathrm{\!_F}^2}{2m_em_\mathrm{eff}}.
\ee
Here, the density of states per plaquette and spin is a constant
\be
\rho(\eF)=\frac{\md N(\eF)}{\md \eF}
=\frac1{2\pi m_\mathrm{eff}(\hbar^2/m_ea_0^2)}=\mathrm{const}.
\ee
In this special case, from \Eqref{Euler-Asynchronism_energy_}
if we double integrate for $\eta<0$ and neglect the contribution of
$\eta>0$ domain, $E_\mathrm{C}\sim\eF$, c.f. Ref.~\cite[Sec.~39]{LL9},
we with up with the three-dimensional result 
$E_\mathrm{C}=0.49~\eF$; the dimensionless factor in front of the Fermi energy 
is irrelevant for qualitative considerations.
For comparison, we reproduce the equation~\cite[Eq.~(39.10)]{LL9} 
\be
\Delta_0=2E_\mathrm{C}\exp(-1/g\nu_\mathrm{F}),
\qquad \nu_\mathrm{F}=mP_\mathrm{F}/2\pi^2\hbar^3,
\ee
where $g$ is the constant in front of a $\delta$-like
attracting interaction between electrons and $\nu_\mathrm{F}$
is the 3-dimensional energy density of electrons at Fermi energy per
spin, unit volume and unit energy.

Here we report also the Pokrovsky equation~\cite{Pokr:61} for the order parameter of anisotropic
superconductors
\begin{subequations}
\begin{align}&
-\ln\frac{\Xi(T)}{\Xi(0)}=2\langle\chi_\mathrm{p}^2F(\Delta_\mathbf{p}(T)/T)\rangle,\\&
F(x)\equiv\int_0^\infty\frac{\md u}{\sqrt{u^2+x^2}\,
[\exp(\sqrt{u^2+x^2})+1]},
\end{align}
\end{subequations}
which gives 
\begin{subequations}
\begin{align}
\frac{2\Delta_\mathrm{max}}{T_c} & =\frac{2\pi}{\gamma}
\tilde\chi_\mathrm{max},\\
\tilde\chi_\mathrm{max} & \equiv
\dfrac{\mathrm{max} \left|\chi_\mathbf{p}\right|}
{\chi_\mathrm{av}} 
=\frac{2\Delta_\mathrm{max}}{
\exp\left(
\langle\Delta_\mathbf{p}^2\ln\Delta_\mathbf{p}^2\rangle/
\langle\Delta_\mathbf{p}^2\rangle
\right)
},
\end{align}
\end{subequations}
where $\tilde\chi_\mathrm{max}$ is the modulus of the rescaled gap
anisotropy that is specific to each
superconductor and is a very informative experimentally accessible
quantity. The Pokrovsky parameter $\tilde\chi_\mathrm{max}$
is an important feature of the theory of anisotropic superconductors 
which describes the deviation of $2\Delta_\mathrm{max}/T_c$
ratio from the isotropic gap BCS value 3.53.
For the model example of constant Fermi velocity
$v_\mathrm{_F}=\mathrm{const}$, in two dimensions
for purely $d$-wave superconductor 
$\chi_\mathbf{p}\propto\cos 2\varphi$ where $\varphi=\arctan(p_x,p_y)$,
we have \cite[Eq.~(3.70)]{MishonovPenev:11}
\be
\tilde\chi_\mathrm{max}^\mathrm{(model)}
=\frac{2}{\sqrt{\e}},\quad
\frac{2\Delta_\mathrm{max}}{T_c}
=\frac{2\pi}{\gamma}\frac{2}{\sqrt{\e}}=4.28.
\ee
Not knowing about the Pokrovsky theory~\cite{Pokr:61} and the integral
\be
\int_0^{\pi/2}\cos^2\!\varphi\,
\ln\!|\!\cos\varphi|\,\md\varphi=\frac{\pi}{8}\ln(\mathrm{e}/4)
\ee
in their numerical analysis of the BCS equation,
Won and Maki~\cite{Won:94} obtained
\be
\tilde\chi_\mathrm{max}^\mathrm{(model)}
=\frac{2}{\sqrt{\e}}\approx1.21306\dots
\ee
We wish also to add that originally this relation was derived
using the $\exp(-1/\lambda)$ as a small parameter but for CuO$_2$
this is an exact result for the Kondo interaction incorporated in the LCAO electron band approximation.

%Now, to complete the description of the calculation.
%As already stated, the value of $J_{sd}$ used for obtaining the result in \Fref{Fig:rTc} is determined from the linear regression from \Eqref{lambdaBCS_}.
%Similarly, the linear regression in \Fref{Fig:lmbTc_} is obtained from the calculation of \Eqref{TcBCS_}.
%These are the main results and achievements of this study.

Let us briefly discuss the set of approximations used in this paper.
First of all, we use the standard BCS scheme which can be derived as 
a trial function approximation of a many body Hamiltonian.
First, it is known that the BCS self-consistent
approximation holds when $T_c$ is much smaller than other energy
parameters of the Hamiltonian, such as the Debye frequency for phonon superconductors 
and $J_{sd}$ in our theory.
Second, the LCAO approximation within 
Complete Neglect of Differential Overlap (CNDO):
used for the momentum dependence of energy bands 
and the shape of CEC.
It is well-known that these approximations 
perfectly reproduce the results of 
LDA calculations and ARPES measurements of Fermi contours.
The third and most important approximation is to take into account only the dominant \textit{s-d} interaction in our analysis 
of $T_{c,\mathrm{max}}$.
Definitely, if we try to calculate the doping dependence of 
$T_c(\tilde p)$,
we have to take into account all other super-exchange amplitudes
$J_{dd}$, $J_{sp}$, $J_{dp}$ along with the electron phonon interaction.
All those interactions are de-pairing for the d-type
superconducting gap in cuprates.
Forth: We suppose that the electronic state of the CuO$_2$ 
plane is space homogeneous.
Which is not true for under-doped cuptrates with a
pseudogap.
Actually, a pseudogap, if any exists, is smaller close to optimal doping and
this justifies this last approximation.

After this brief review of the analytical results for the $J_{sd}$-LCAO
theory of the CuO$_2$ superconductivity, we can address the problem of
calculation of $T_c$ versus Fermi surface shape correlation.

\subsection{LCAO formulas for the Fermi Contour}

%After the review of the analytical formulae, we address
%their application for experimental data processing.
At known energy dependence of the coefficients
in the secular equation for the energy spectrum,
we can express the dimensionless ratio
\begin{align}
\frac{t^\prime}{t}&
=\dfrac{1}{2+4\,\dfrac{\mathcal{B}(\eF)}{\mathcal{A}(\eF)}}
=\dfrac{1}{2+4\,\dfrac{\mathcal{B}_f}{\mathcal{A}_f}}
\equiv\mathcal{F},
\label{t'/t_}
\end{align}
determining the shape of CEC and the Fermi contour at $\epsilon=\eF$.
\Eqref{t^prime_t} can be rewritten as
\be
\cos p_x+\cos p_y-2\mathcal{F}\cos p_x \cos p_y =\mathrm{const},
\ee
because the parameters $t$ and $t^\prime$ have no direct physical
meaning
in the real electronic structure of the CuO$_2$ planes.
The hole pocket contour passes through the points
$\tilde{\mathcal{D}}= (p_d,\,p_d)$ and
$\tilde{\mathcal{C}} =(\pi,\,p_c)$
Hence, we introduce 
\begin{subequations}
\begin{align}
& x_d=(-\mathcal{B}+\sqrt{\mathcal{B}^2-\mathcal{A}\mathcal{C}})/\mathcal{A}
=\sin^2(p_d/2),\\
& x_c=y_c=-(\mathcal{B}+\mathcal{C})/(\mathcal{A}+\mathcal{B})=\sin^2(p_c/2).
\end{align}
\end{subequations}
Introduced in such a graphical manner,
the parameters $x_c$ and $x_d$ can be used to fit
CEC to ARPES data. We introduce the results from this fit
\begin{subequations}
\label{CEC_fit_}
\begin{align}
& \mathcal{A}_f=2x_d-x_c-1,  \\ %& x_d=\sin^2(p_d/2),\\
& \mathcal{B}_f=x_c -x_d^2,  \\ %& x_c=\sin^2(p_c/2),\\
& \mathcal{C}_f=x_d^2(x_c+1)-2x_c x_d, \\
& \mathcal{A}_f \, xy+\mathcal{B}_f \, (x+y)+\mathcal{C}_f =0, \\
& \mathcal{A}_f /\mathcal{B}_f = \mathcal{A} /\mathcal{B}.
\end{align}
\end{subequations}
and the shape parameter $t^\prime/t$ remains unchanged.
The subscript $f$ in \Eqref{t'/t_} refers to a fit
of the points 
$(p_d,\,p_d)$ and $(\pi,\,p_c)$ of known Fermi contours from ARPES experimental data or electron band calculations, while the argument $\eF$ denotes the Fermi energy from the filling factor $f_h$ at fixed other LCAO parameters.

Now we can determine the parameters of the Hamiltonian.
For single site energies and hopping integrals, we start with the set
of LCAO-parameters given in Ref.~\cite{Pavarini:01}.

It is well-known that electron band calculations
systematically give significantly broader band than their
counterpart extracted from ARPES data.
To surmount this discrepancy, we perform a renormalization of all energy parameters $\epsilon_s$, $\epsilon_p$, $\epsilon_d$, 
$t_{pd}$, $t_{sp}$, and $t_{pp}$ with a common divider $Z = 1.37$
to obtain approximately equal theoretically calculated Fermi velocity 
$V_\mathrm{_F}=\partial\epsilon/\partial P$
and experimentally evaluated
$V_\mathrm{_F} \approx 1.25\ \text{eV\! \r{A}}/\hbar \approx 190$~km/s for Bi$_2$Sr$_2$Ca$_1$Cu$_2$O$_8$
along the nodal direction $(0,\,0)$-$(\pi,\, \pi)$.
In this rough evaluation, we have taken the
lowest slope of the dispersion curve in the middle of the slope in
Fig. 1e. of Ref.~\cite{Johnson:01},
cf.~\cite[Eqs.~(65.12-13)]{LL9}
\begin{align}
& \varepsilon-\eF\approx v_\mathrm{F}^{(0)}
(p-p_\mathrm{F}), \quad\mbox{for  } 
\varepsilon-\eF\gg E_\mathrm{D}\equiv\hbar\omega_\mathrm{D}, \nn \\
& \varepsilon-\eF\approx v_\mathrm{F} 
(p-p_\mathrm{F}), \quad\mbox{    for  } \nn
\varepsilon-\eF\ll E_\mathrm{D},
\end{align}
we consider the region $|\epsilon-\eF|\ll\hbar\omega_\mathrm{D}$.
In these formulas $v_\mathrm{F}^{(0)}$ corresponds to Fermi 
velocity of a purely electronic system far from $\eF$,
while $v_\mathrm{F}$ includes the influence on the Fermi velocity of the lattice polarization.
For our calculation,
the re-normalized numerical values of
the LCAO parameters
$\epsilon_i\rightarrow\epsilon_i/Z$ and $t_{ij}\rightarrow t_{ij}/Z$
are listed in  Table~\ref{tbl:in_energy}.
%%%%%%%
\begin{table}[ht]
\caption{
Single site energies $\epsilon$ and hopping amplitudes $t$ in eV.
The values are taken to be approximate to the ones from
Refs.~\onlinecite{Andersen:95} and \onlinecite{Pavarini:01}.
}
%\begin{tabular}{ >{\centering}p{0.7cm} >{\centering}p{0.7cm} >{\centering}p{0.7cm}
%				      >{\centering}p{0.7cm} >{\centering}p{0.7cm} >{\centering}p{0.7cm} 
\begin{tabular}{ccccccccc}
\hline \hline
\ \ $\epsilon_s$ \ \ & \ \ $\epsilon_p$ \ \ & \ \ $\epsilon_d$ \ \ &
	\ \ $t_{sp}$ \ \  & \ \ $t_{pp}$\cite{Mishonov:96} \ \ & \ \
	$t_{pd}$ \ \ & \ \ $f_h$ \ \ & \ \ $a_0$ \ \ & \ \ $T_{c,\,\mathrm{max}}$ \\ 
			&  \\ [-1em] 
			4.0	& -0.9 & 0.0 &	2.0	& 0.2 & 1.5 & 0.58 &3.6~\r{A}& 90~K \\
\hline \hline
\end{tabular}
\label{tbl:in_energy}
\end{table}
%%%%%%%
If the comparison between ARPES data and \textit{ab initio} band calculation requires more significant renormalization, 
that will lead to increase of the effective masses and decrease of the $J_{sd}$ exchange amplitude.
In the introduction we stated that the electron band theory is well-applicable to the cuprates. 
But this is an oversimplification, since for highly
correlated heavy fermion systems, band theory using the Local Density Approximation (LDA) 
e. g. the Wien2K code gives a good description of the
shape of the Fermi surface.
For a review the interested reader may consult Ref.
\cite{Hewson:93} containing extensive discussions on the \emph{s-d} model.
On the other hand, it is well-known that LDA does not account correctly for the 
Fermi velocities or effective masses. 
For example, for the single layer compound 
Tl$_2$Ba$_2$CuO$_{6+\delta}$ (Tl:2201) where
quantum oscillations from a large Fermi surface are observed, 
the value of the effective mass $m_\mathrm{eff}$
is 3 times larger than the predicted by LDA theory \cite{Rourke:10}. 
For the material Tl$_2$Ba$_2$CuO$_{6+\delta}$ there is a simple LCAO approximation
for 3D Fermi surface~\cite{MishStoev:10,PhysScr}.
Moreover, results the specific heat and the static susceptibility data 
for other cuprates Ref.~\cite{Cooper:22}, 
gives similar values of $m^*$ and 
a Wilson ratio of approximately $R_w\approx 1.3 $.
For an extended discussion of the importance of $R_w$
see Ref.~\cite{Hewson:93}.

We assume that the introduced energy renormalization
of the 4-band electronic structure
\be
\epsilon\rightarrow\epsilon/Z_\epsilon,\qquad
t_{ab}\rightarrow t_{ab}/Z_\epsilon
\ee
is an admissible approximation.
The ARPES data plotted in \Fref{Fig:rTc} suggest how
the energy renormalization divider $Z_\epsilon$ can be determined.
Moreover, within the LCAO approximation
the analytical result 
for the Fermi velocity \Eqref{velocity_} can be used for the purpose of comparison
with ARPES data for overdoped cuprates
whose whole Fermi contour can be constructed.
From the LCAO approximation of the band structure, we can evaluate $m_\mathrm{eff}^\mathrm{(LDA)}$.
Another possibility follows from
de Haas-van Alphen effect study of the overdoped cuprate Tl$_2$Ba$_2$CuO$_{6+\delta}$ \cite{Rourke:10}.
Here we recall the textbook example that for 
highly anisotropic almost two dimensional (2D) materials 
with a single pocket for the Fermi contour all definitions
for the effective masses coincides:
the optical mass~\cite[Eq.~(3.01)]{MishonovPenev:11}, 
the cyclotron mass~\cite[Eq.~(57.6)]{LL9}, and the masses 
$m_\mathrm{eff}$ related to
the density of states and heat capacity.
However, ARPES and de Haas-van Alphen effect studies cannot be
performed on all cuprates that are relatively clean materials with
electron effective mass $m_\mathrm{eff}=\frac12 m^*$ with $m^*$ the mass
of the super-fluid charge carriers
$m^{*}$~\cite{MishonovPenev:11};
see also Refs.~\cite{MishonovZahariev:99,PhysScr}.
We can determine a simple prescription for the energy renormalizing
parameter
\be
Z_\epsilon=\frac{m^*/2}{m_\mathrm{eff}^\mathrm{(LDA)}},\qquad
Z_\epsilon-1\gg R_w-1.
\ee
Till then the density of charge carriers can be evaluated 
by chemical means and the mass can be estimated
through the penetration depth.
Indispensably this renormalization will reduce
$J_{sd}$, which now seems too large.
In any case the speculated energy renormalization 
will not alter the correlation coefficient of the linear regression
depicted in \Fref{Fig:lmbTc_} and \Fref{Fig:tpt-lambda_}
This linear dependence is an immanent property of the 
\textit{s-d} pairing mechanism.
In a completely different physical picture, the exchange interaction
gives the Kadowaki-Woods~\cite{Kadowaki:86} 
% K.~Kadowaki and S.~B. Woods, Solid State Comm, 
% \textbf{58}, 507 (1986).
relation $A_{_\mathrm{BLP}}\propto \gamma_C^2$, where
$\gamma_C$ is the coefficient of the low-temperature linear heat capacity $C\approx
\gamma_C T$ and $A_{_\mathrm{BLP}}$
the coefficient of the low-temperature resistivity $\varrho=A_{_\mathrm{BLP}}T^2$.
The similarities between heavy fermion compounds 
and high-$T_c$ cuprates are discussed in
Ref.~\cite{Hewson:93}.

The difference between experimentally observed masses and 
effective masses evaluated by LDA is actually a very old problem.
When for the first time the effective mass of cooper pair was determined
by electrostatic doping of 
thin-YBa$2$Cu$_3$O$_{7-\delta}$-film-kinetic 
inductance 
it was reported~\cite{Comment:91} 
that $m^*/2\approx 5.5\,m_e$.
The recent estimate~\cite{Rourke:10} obtained
by magnetic oscillations of de Haas-van Alphen effect confirms this
result $m_\mathrm{eff}\approx 5.2\,m_e$
for Tl$_2$Ba$_2$CuO$_{6+\delta}$.
In Table~\ref{tbl:out_energy_} the optical mass 
$m_\mathrm{opt}=1.22\,m_e$
calculated by LCAO approximation of the LDA from
Ref.~\cite{Pavarini:01} is given.
Thus, for illustration, we obtain another evaluation
for the energy renormalization denominator
$Z_\epsilon=\frac{5.2}{1.22} \approx 4.2 $.
To summarize, the energy
renormalization allows for
the possibility band energies calculated by LDA to be used
as parameters in low energy effects close to the Fermi level.
The energy renormalization constant $Z_\epsilon$
provides the possibility
to preserve the shape of the Fermi contour determined by the 
position of the Cu$4s$ level and simultaneously to fit the effective
mass, an important quantity of the BCS theory related to the density
of states close the Fermi level.
The purpose is to determine the parameters of 4-band LCAO
model which allows to calculate all properties of
superconductivity like temperature dependencies of 
the superconducting gap, penetration dept, heat capacity,
condensation energy $B_c^2(T)/2\mu_0$, etc.
A single CuO$_2$ plane of Tl$_2$Ba$_2$CuO$_{6+\delta}$
could be the best material to test the initial set of parameters.
For this clean material with relatively low $T_c$, perhaps even the cyclotron mass can be directly determined by cyclotron resonance.
Later, one can calculate as a perturbation the influence of
double or triple layers, apex oxygen chains etc.

In Table~\ref{tbl:out_energy_}  the calculated Fermi energy for the
optimal doping $\eF$, the energy of the uppermost energy
level of the conduction band, the Van Hove energy
$\epsilon_\mathrm{_X}=\epsilon_{0,\pi}$,
the calculated according to \Eqref{Euler-Asynchronism_energy_}
Euler-Mascheroni energy $E_\mathrm{C}$
with small parameter $\epsilon=1\,\mu$eV
and other parameters of the theory are given.
%$\epsilon_\mathrm{_M}=\epsilon_{\pi,\pi}$,
%%%%%%%
\begin{table}[ht]
\caption{Output parameters of our numerical calculation, the extra numbers are only for a numerical test.
The new quantities are the values of the \textit{s-d} exchange amplitude $J_{sd}$ and the effective masses derived from the parameters of electron band calculations~\cite{Pavarini:01}. 
The value of $\tilde{\chi}_\mathrm{max}=1.167$ is within  5\% accuracy 
of the model evaluation for a pure \textit{d}-wave gap with isotropic Fermi velocity which gives 
$\tilde\chi_\mathrm{max}^\mathrm{(model)}=2/\sqrt{\e}=1.213$.}
\begin{tabular}{ r r r }	
\hline \hline
$E_\mathrm{C}$ = 1.403~eV & \ \ $\lambda$ = 0.188         \ \   & $m_\mathrm{top}$ =  1.15 \\[0.1cm]
$\eF$ = 1.351~eV          & $\tilde{\chi}_\mathrm{max}$ = 1.167 & $m_c$ =  1.28            \\[0.1cm]
$\epsilon_{\mathrm{_M}}$ = 3.061~eV  & $\langle \chi^2 \rangle$ = 0.044 & $m_\mathrm{opt}$ = 1.22 \\[0.1cm]
$\epsilon_{\mathrm{_X}}$ = 0.851~eV  & $\langle \chi^2 \rangle^2/\langle \chi^4 \rangle$ = 0.737 & $r$ = 0.365~eV \\[0.1cm]
$E_0$ = 0.528~eV  & $\rho_\mathrm{_F} = 0.385~\mathrm{eV}^{-1}$	&  $2/\sqrt{\e}$ = 1.213\\[0.1cm]
$J_{sd}$ = 5.593~eV & \ \ \ $2 \Delta_\mathrm{max}/T_{c,\,\mathrm{max}}$ = 4.116 & \ \ \ $V_0$ = 0.488~eV \\[0.1cm]
\hline \hline
\end{tabular}
\label{tbl:out_energy_}
\end{table}
%%%%%%%

Then, we have the appropriate value $t^\prime/t=0.542$
for a $90$~K cuprate and the calculation fixes $J_{sd}$ from \Eqref{BCS_gap_equation_},
with $\Xi=0$.
Now that all the parameters of the Hamiltonian are determined, we can
predict the experimental results by calculating all
relevant to the BCS theory quantities.
Varying $\epsilon_s$, we calculate $t^\prime/t$ and $\lambda$
according to \Eqref{t'/t_} and \Eqref{lambdaBCS_}, respectively.
Let us point out that the product entering expression \eqref{lambdaBCS_} of the
BCS coupling constant $\lambda$
is a Fermi contour integral
\be
\langle \chi^2\rangle\rho_\mathrm{_F}
=\frac{1}{(2\pi)^2}\oint\frac{\md p_l}{v_\mathrm{_F}}
\chi_\mathbf{p}^2
=\frac{8}{(2\pi)^2}\int_{p_d}^\pi
\frac{\md p_l}{\md p_x}\frac{\chi_\mathbf{p}^2}{v_\mathrm{_F}}
\,\md p_x
\label{chi2rho}
\ee
with a complex integrand $\chi$ given by \Eqref{chi_analytical_}.
Varying the Cu4$s$ energy level $\epsilon_s$,
we calculate $\lambda(\epsilon_s)$
and $t^\prime/t(\epsilon_s)$ at fixed all other parameters.
The result of the ensuing $t^\prime/t (\lambda)$ curve is 
shown in \Fref{Fig:tpt-lambda_}.
This seemingly noninteresting straight line (within the accuracy of the numerical calculation) represents the relation between the BCS coupling constant 
$\lambda$ defined in \Eqref{lambdaBCS_}
with $J_{sd}$ given in Table~\ref{tbl:out_energy_}
and the ratio of the tight-binding parameters $t^\prime/t$ calculated
with the aid of \Eqref{t'/t_} and \Eqref{ABC_}.
According to \Eqref{TcBCS_}, $\lambda$
has the main influence on the critical temperature $T_c$.
The intricate integral representing
$\langle\chi^2\rangle\rho_\mathrm{_F}$ 
with the analytic expression \Eqref{chi_analytical_} substituted
in the Fermi contour, after averaging \Eqref{velocity_},
see also \Eqref{chi2rho},
gives a little hope for an analytical solution.
for both variables we have derived the subtle
explicit expressions \Eqref{t'/t_} and \Eqref{TcBCS_} within the used LCAO approximation for the electronic band structure.
Thus, we end up with almost a linear dependence between the variable related to the interaction $\langle\chi^2\rangle\rho_\mathrm{_F}$ 
and the parameter $t^\prime/t$ that defines the shape of the Fermi contour.
Notice, that relations
attained relying on a pure phenomenological intuition~\cite{Pavarini:01}
finds a standard BCS interpretation using the Kondo pairing
interaction and the LCAO approximation of the independent electronic band function.

\begin{figure}[ht]
\centering
\includegraphics[scale=0.5]{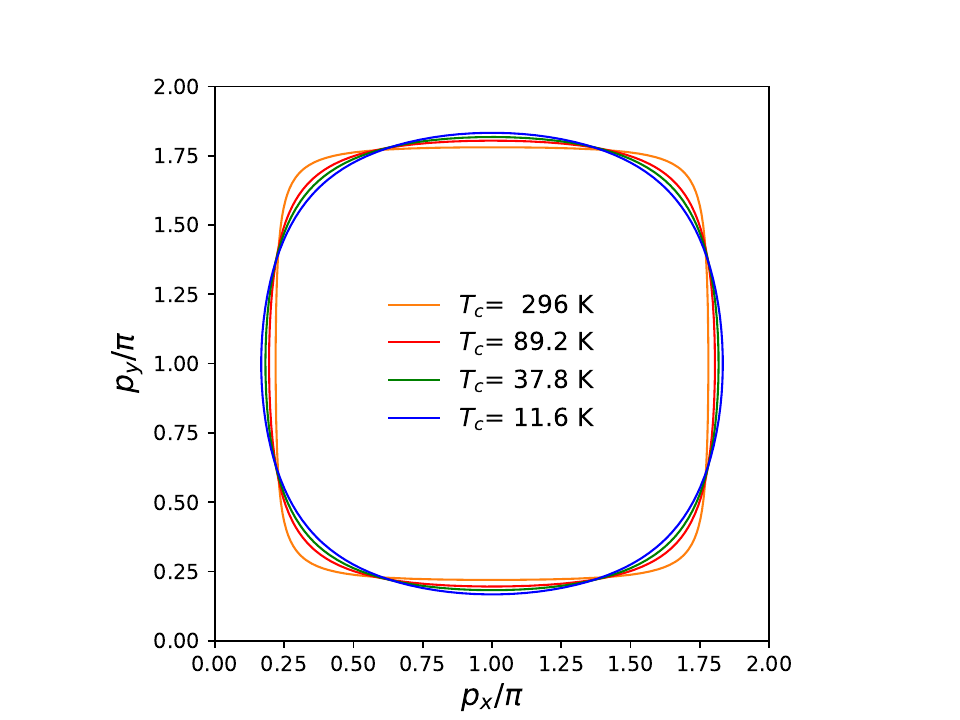}
\caption{
Several constant energy curves (CEC) corresponding to 
Fermi contours with optimal doping according to
\Eqref{t^prime_t} and \Eqref{spectrum_}.
The area of the hole pocket is 58\% from the Brillouin zone.
The curves corresponds to different ratios $t^\prime/t$
and different $T_{c,\mathrm{max}}$ 
according to the curve from  
\Fref{Fig:rTc}.
More rounded edges correspond to lower critical temperatures.
The almost square-edged contour at the Brillouin zone diagonals does not correspond to a real cuprate.
It is just an extrapolation of the case when 
Cu4$s$ level $\epsilon_s$ is only slightly 
above the Fermi level $\eF$.
In order to reach room $T_c$
(blue dream of physicists) with a cuprate is necessary to create
a meta-stable layered structure in which
an appropriate apex ligand decrease maximally the Cu4$s$ level.
Perhaps it is a doable task for MBE 
(Molecular Beam Epitaxy) technologists.
}
\label{Fig:CECs}
\end{figure}

\begin{figure}[ht!]
\centering
\includegraphics[scale=0.55]{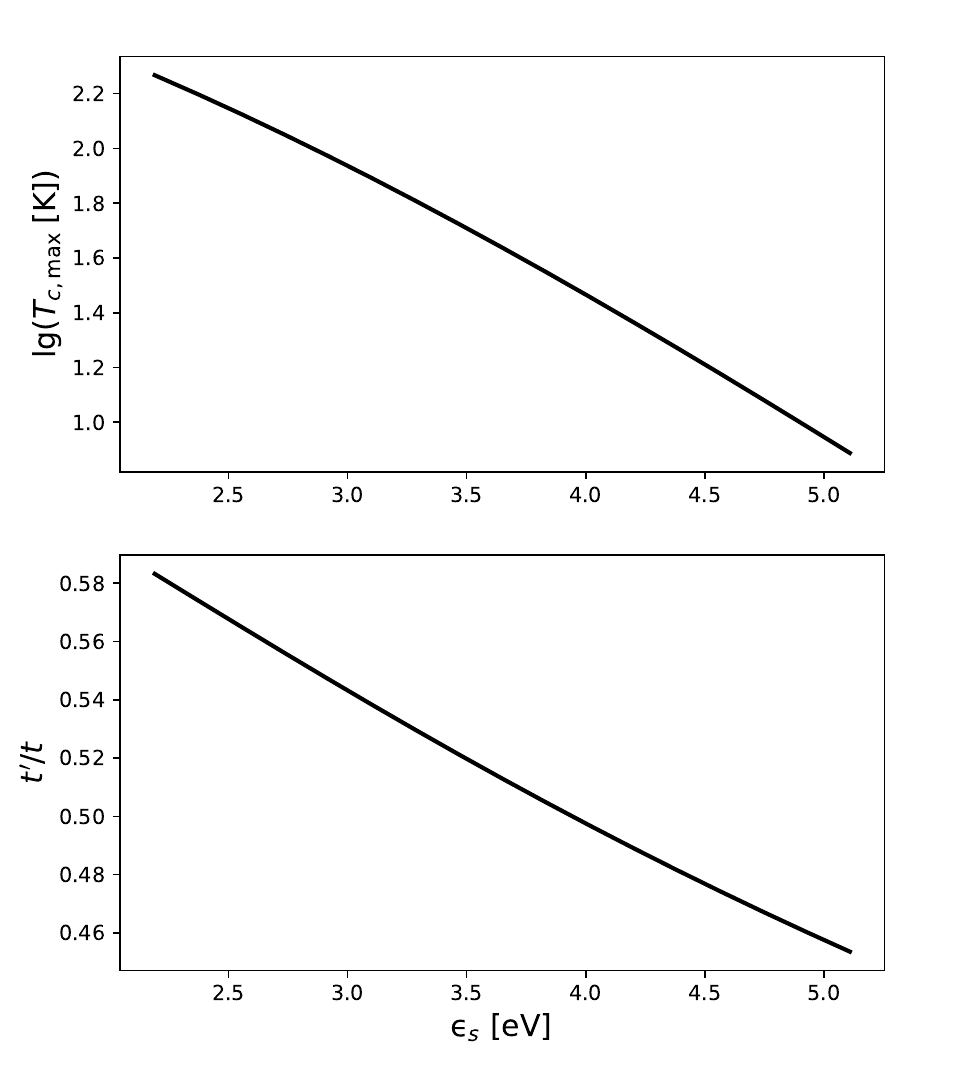}
\caption{
(Up) R\"{o}hler~plot, the logarithm of the maximal critical temperature 
$T_{c,\mathrm{max}}$ versus the energy difference $\epsilon_s$
between the  Cu4$s$ empty band and the conduction Cu3$d_{x^2-y^2}$ band 
at the $\Gamma$ point
$p_x=p_y=0$. 
Calculations are numerical solution of the BCS equation
\Eqref{BCS_gap_equation_}.
(Down) Andersen plot, the dimensionless ratio $t^\prime/t$ as a function of  $\epsilon_s$ at fixed other LCAO and $J_{sd}$ parameters. 
Calculations are performed using \Eqref{t'/t_}.
The almost linear dependencies are consequences of 
the energy denominators of the perturbation 
theory~\cite{BJP:11}.
}
\label{Fig:Roeler_plot}
\end{figure}

The results of the numerical calculations reveal the almost linear dependence between
$\mathcal{Y}(\epsilon_s)\equiv\langle \chi^2\rangle\rho_\mathrm{_F}$ and 
$\mathcal{X}(\epsilon_s)\equiv t^\prime/t$.
The quasi proportionality $\mathcal{Y}\propto \mathcal{X}$, when only
$\epsilon_s$ is varied is a highly nontrivial result which leads to
the final explanation of the correlation in \Fref{Fig:lmbTc_} (see also the
linear dependence in \Fref{Fig:tpt-lambda_}).
According to the BCS result for the critical temperature
\Eqref{TcBCS_}, we observe a direct relation between 
$\ln T_{c,\,\mathrm{max}}$ and the reciprocal BCS coupling constant
$1/\lambda$,
which is determined mainly by the relative position of the Cu4$s$ level
with respect to Cu3$d_{x^2-y^2}$.
Within the tight-binding modeling by Honerkamp and
Rice~\cite{Honerkamp:03} and Sarasua~\cite{Sarasua:11} it was found
that increasing the ratio $t^\prime/t$ favors electron pairing.
In the present study we reveal that this ratio is determined by the Cu4$s$ level.
As an illustration several Fermi contours corresponding to the
optimal doping $\tilde p=16\%$ are depicted in
\Fref{Fig:CECs}. 
%%%%%%%%%%%
%%%%%%%%
The correlation between the shape of the Fermi contour and the maximal critical temperature observed by Andersen has been confirmed by R\"{o}hler who observed the correlation between the critical temperature and the position of the Cu4$s$ level. 
The corresponding R\"{o}hler-Andersen plots are depicted in \Fref{Fig:Roeler_plot}. 
%%%%%%%%%%%
%%%%%%%%
Unfortunately, no
chemical methods allowing for ligand engineering to further decrease
of $\epsilon_s$ are available.

To avoid any misunderstanding, we clarify that $T_{c,\,\mathrm{max}}$ depends also on doping or chemical potential.
But the named correlations are valid solely for optimally doped
cuprates with relatively similar area of the hole pocket. Let us
recall that this optimal doping fixes the chemical potential or Fermi level~\cite{Pavarini:01}.
Moreover, interesting physics of underdoped cuprates is often
irrelevant for optimal doping.
However, high-$T_c$ cuprates have attracted great attention due to
their high critical temperature and the purpose of our study is to
reveal what is the pairing mechanism leading to such a high critical
temperature at optimal doping.

\section{Conclusion}

\subsection{The beginning of the emerging consensus}
In eqarly studies \cite{Micnas:90}, the calculated bandwidths were much higher than the
experimentally evaluated and, in this sense, high-$T_c$ cuprates were considered as narrow-band systems with local non-retarded attractive interactions.
Within the BCS scheme, the authors of Ref. conclude that in
narrow-band systems, electrons can interact with each other via a
short-range non-retarded attractive potential. They did not
exclude that it could be a purely electronic mechanism.
They considered multiple limiting scenarios,
such as the extended Hubbard model with negative $U$.
and suggested that the spin exchange 
may be regarded as a correlated hopping.
In a way, the present study was provoked by
restrictions set in Ref.~\cite{Micnas:90}, namely what interaction in
condensed matter theory satisfies the requirements:
(1) short-range, 
(2) non-retarded, 
(3) attractive, i.e. described by the extended negative-$U$ Hubbard model, 
and (4) based on spin exchange and correlated hopping?
To the best of our knowledge, the Kondo $s$-$d$ exchange interaction 
is the only candidate \cite{Micnas:90} to do so.

In the present study, we advance that cuprate superconductivity
is not a phenomenon related to a single compound 
whose various parameters can lead to
strange properties.
We have several CuO$_2$ plane compounds and
definitely the high-$T_c$ cuprate superconductivity should be
induced by an intensive interaction omnipresent for all transition metal ions. 
Actually, Zener introduced the $s$-$d$ exchange 
long before the BCS theory, and Kondo concluded
that to treat correctly the kinetics of the normal metallic phase demonstrates that
exchange amplitude must be antiferromagnetic.
Incorporating the Kondo interaction in the BCS scheme shows that $J_{sd}$ is equivalent
to a negative-$U$-Hubbard model, but with an important byproduct:
\textit{the analytical formula for the $d$-wave gap anisotropy}.
Recently the development of physics of cuprates,
doped bismutates and iron-based systems
boosted many contemporary theoretical works on 
the extended Hubbard model with pair 
hopping~\cite{Kapcia:12} and local electron 
pairing~\cite{Kapcia:13}.

\subsection{General discussion}
Let us first introduce the approximations used in this study:
(1) the BCS approximation applicable when $T_c$
is much smaller than bandwidth of conduction band, 
and 
(2) the LCAO approximation for the electron band structure.
Within these approximations exact analytical formulas are well-known and 
in order to explain the experimentally observed correlation,
the corresponding equations are numerically solved.

Comparing the results for the Kondo interaction with the results of the phonon model,
see for example the recent study by Marsiglio~\cite{Marsiglio:18},
our \Eqref{lambdaBCS_} and \Fref{Fig:lmbTc_} are very similar
to \cite[Eq.~(26), Fig.~1 and Fig.~4]{Marsiglio:18}.
However, every model for cuprate superconductivity has its own
intrinsic parameters that are difficult to calculate using \textit{ab initio}. 
For example, the $J_{sd}$ exchange constant and the electron phonon coupling constant.
According to Ayres, Katsnelson and Hussey~\cite{Ayres:22} the single band Hubbard model combined with density functional theory is the minimal model for studying the physics of cuprates.

When the parameters inherent to a theory are determined via a fit to
experimental data, the approximations used to obtain the dimensionless
parameters have to be verified. In our case for the applicability of the
BCS approach, we have to verify whether $T_{c,\,\mathrm{max}}\ll
E_\mathrm{C}$ holds.
In general, there are numerous models intended to describe CuO$_2$
superconductivity.
Thus, an important question naturally arises: \textit{How to determine the parameters of
these effective Hamiltonians?} Moreover, how to derive microscopically
the influence of the Cu4$s$ level on the relevant parameters?
Without incorporating the Cu4$s$ orbital, it is highly unlikely to
obtain adequate description of electronic processes in high-$T_c$
cuprates, let us note that such attempts are still being conducted nowadays~\cite{OMahony:22}.
The superconducting phase transition is determined by the pairing
interaction and if the position of the Cu4$s$ level has a large
influence on $T_{c,\,\mathrm{max}}$ this is a hint that the Cu4$s$
state is crucial to the pairing interaction,
in our case this is \Eqref{Eq.Kondo}.

In the The last few decades, it has been revealed that high-$T_c$ 
materials possess all the properties of BCS superconductors: 
charge of Cooper pairs, band structure, superconducting gap, etc.
For the Kondo interaction applied to the CuO$_2$ plane the pairing
function is factorizable.
For his theory of anisotropic gap superconductors, 
Pokrovsky derived the factorizable pairing interaction 
within the weak coupling approximation ($\lambda\ll1$),
but this condition is not necessary for cuprates.
In other words, the Pokrovsky theory for anisotropic gap superconductors
is applicable for the exchange pairing in cuprates even for moderate coupling constants, say $\lambda\sim1/2$.
However, the main result of the present study is not the BCS estimation of $T_{c,\,\mathrm{max}}$, 
but rather the fact that that only the Kondo \textit{s-d} interaction 
considered as a pairing mechanism explains the well-known  experimentally observed correlation 
between the critical temperature of optimally doped cuprates 
and the shape of the Fermi contour.

In the long-term search of a suitable high-$T_c$ pairing mechanism,
the correlation of $T_c$ with all properties of 
high-$T_c$ superconductors was carefully studied.
The result is that the most pronounced correlation
is the one between $T_{c,\mathrm{max}}$ and the shape of the 
Fermi contour parameterized by a single dimensionless
parameter $t^\prime/t$.
But the Fermi surface is a property of the
independent non-interacting electron approximation.
Thus, the named correlation can only reveal the hidden influence of
a strong interaction. Moreover, the shape of the Fermi contour could only be related
to the matrix elements of the pairing interaction.
Whence, our study reveals that the Kondo $s$-$d$ interaction is
the predominant mechanism explaining this correlation so far and it
will persist until a theory with better or at least similar
explanation of the same correlation emerges.

The above analysis of the correlation depicted in \Fref{Fig:lmbTc_},
reveals that the BCS pairing theory within an acceptable
accuracy qualitatively is even able to explain the main property of the HTS cuprates --
their high critical temperature $T_{c,\,\mathrm{max}}$ at optimal doping.
%To repeat, due to lack of alternative explanation cf.  
%Ayres, Katsnelson and Hussey~\cite{Ayres:22}
%we arrive at the conclusion that the long
%sought mechanism has already been found --
%the well-known Kondo exchange interaction
%applied to the conduction band charge carriers.
%Roughly speaking, science begins with simplicity.
The significant correlation coefficient in the linear regression
depicted in \Fref{Fig:lmbTc_} suggests
that the correlation introduced in Ref.~\cite{Pavarini:01} is
a correlation between $T_c$ and the dimensionless
coupling constant $\lambda$ supposing in a very rough 
initial approximation that $J_{sd}$ depends weakly on the studied
material.
In any case, the Kondo interaction as a pairing mechanism
in cuprates deserves a place on the arena
together with all other mechanisms attempting to explain
the correlation of $T_c$ and other properties of 
high-$T_c$ superconductors.

\subsection{A historical perspective}
\begin{figure}[ht]
\centering
\includegraphics[scale=0.5]{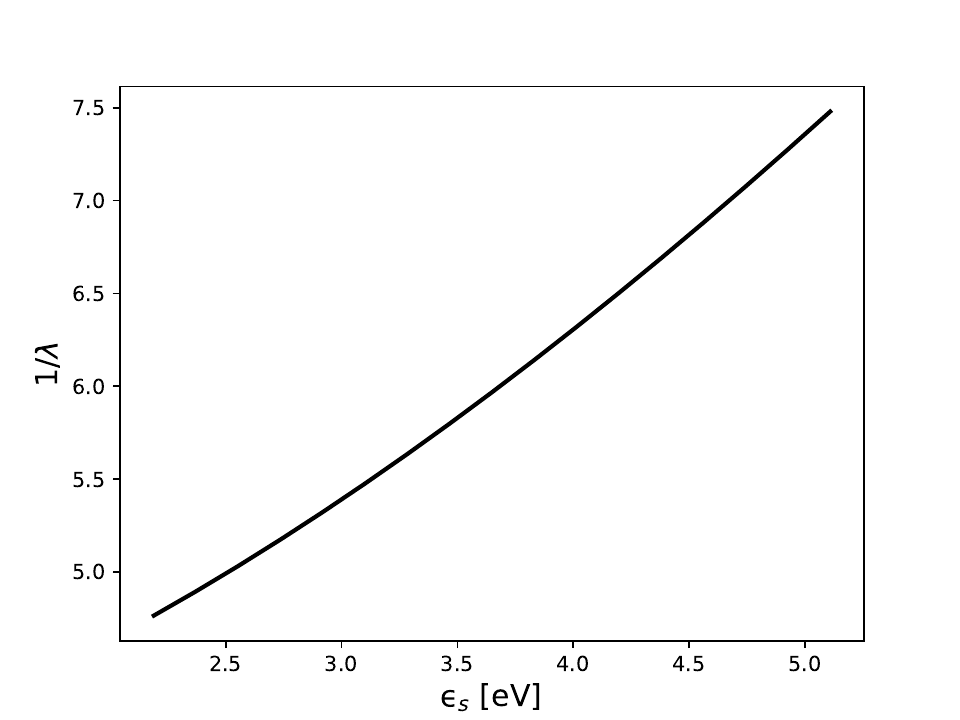}
\caption{Relation between reciprocal value of the BCS coupling constant 
$1/\lambda$ (in ordinate) for the Kondo interaction and $\epsilon_s$ (in the abscissa) 
the position of the Cu4$s$ level from the Cu3$d_{x^2-y^2}$ level.
The almost linear dependence is a consequence of 
applicability of perturbation theory in the band structure~\cite{BJP:11}
}
\label{Fig:lambda-epsilon}
\end{figure}

R\"{o}hler et al.~\cite{Roehler:00,Roehler:00a} 
were the first to mention that the position of the Cu4$s$ level
and the Cu4$s$ hybridization are related to the increase of the
critical temperature.
They reported the correlation between $T_c$ and $\epsilon_s$ analyzing
only a few experimental points. In the framework of our theory 
this correlation has a natural explanation.
The dimensionless coupling constant strongly depends on the position
of the Cu4$s$ level as it is depicted in \Fref{Fig:lambda-epsilon}.
%%%%%%%%%%%
%%%%%%%%
According to the BCS formula for the critical temperature
\Eqref{TcBCS_} $-\ln(T_c)$ is a linear function of
$1/\lambda$; the influence of $\epsilon_s$
on $E_\mathrm{C}$ is much smaller.
Hence, one can trace back the influence of $\epsilon_s$ to $\lambda$,
$T_c$ and the Fermi contour shape parameter $t^\prime/t$.
%It is a very challenging theoretical  task to find an alternative explanation of these experimental correlations.
The observation by R\"{o}hler~\cite{Roehler:00,Roehler:00a} 
explains easily why high-$T_c$ cuprates
have so different values of $T_{c,\mathrm{max}}$,
while having one and the same CuO$_2$ plane.
The apex oxygen ligand or the adjacent 
CuO$_2$ plane decreases $\epsilon_s$
by the covalent hybridization.
It is an open technological task 
to develop superconducting electronics
aiming at creating a meta-stable layered structure
with a Cu4$s$ level significantly lowered
by an ingeniously constructed apex ligand.

Later, it was observed that
the 3$d$-to-4$s$-by-2$p$ is the most likely route to high-$T_c$ 
superconductivity~\cite{MishIndPen:02}
and it was concluded that cuprate superconductivity is the new face of the ancestral $s$-$d$ exchange interaction~\cite{MishIndPen:03}.
In this respect, the apex oxygen decreases
$\epsilon_s$ due to hybridization
of the Cu4$s$ with the apex O2$p_z$ orbitals.
On the other hand, the critical temperature of double and triple
CuO$_2$ planes is higher due to the hybridization of adjacent Cu4$s$
ions that decrease $\epsilon_s$, i.e. the energy difference between
Cu4$s$ and Cu3$d_{x^2-y^2}$ levels.
This observation can even lead to the prediction:
\textit{we can increase $T_c$ if in a metastable structure of an apex
ligand can decrease $\epsilon_s$ by a strong hybridization.}
Let us emphasize that the parameter $\left<\chi^2\right>$ describes the
averaged along the Fermi contour probability for one
electron to be simultaneously in the Cu4$s$ and the Cu3$d_{x^2-y^2}$
states.
This hybridization parameter determines $T_c$ and this is the strongest argument that 
the Cu4$s$--Cu3$d_{x^2-y^2}$ Kondo exchange interaction is the
\textit{right} pairing mechanism.

The linear dependence 
in the $\ln T_c$ versus $1/\lambda$ plot suggests that the adequate
variables are properly taken into account.
The original plot $T_{c,\mathrm{max}}$ versus $t^\prime/t$ has gained
broad recognition:
\textit{The matrix elements of the pairing interaction are so closely related
to the shape of the Fermi surface.}
Even if we believe that the Kondo interaction is the long sought
pairing mechanism, it would be a significant success to corroborate
this finding by computing the properties of the normal phase:
the anisotropy of the hot-cold phenomenology of the width of ARPES lines and the linear dependence of the Ohmic resistivity of the normal phase, cf. \cite{MishIndPen:03}.
We hope that the LCAO approach can be used to study the metal-insulator transition,
spin-density waves and their gap-like influence on the electron spectra~\cite{Abrikosov:03}.

In the great period of high-$T_c$ physics practically all processes in condensed matter 
physics were considered as potential pairing mechanisms.
As the Cu ion is at the end of the sequence of transition metals of the Fe group,
the first task is to check whether the most usual $s$-$d$ exchange has the leading contribution as it is for magnetic properties of the iron group compounds and Kondo alloys.

Let us outline briefly the connection the main quantities of the represented theory of superconductivity of cuprate high-$T_c$ superconductors:
\begin{enumerate}
%\begin{itemize}
% 1) 
\item R\"{o}hler~\cite{Roehler:00,Roehler:00a} emphasized
that the hybridization of Cu4$s$ with a conduction band leads
to increasing of $T_c$.
% 2)
\item Pavarini \textit{et al.}~\cite{Pavarini:01} observed
the celebrated correlation between $T_{c,\,\mathrm{max}}$
and the shape of the Fermi contour parameterized by $t^\prime/t$. 
% 3) 
\item The LCAO analysis is similar to the given by Andersen \textit{et
al.}~\cite{Andersen:95,Andersen:96} reveals that the position of
Cu4$s$ is directly related to the $t^\prime/t$ parameter for optimally doped superconductors.
% 4)
\item The critical temperature $T_c$ is much smaller than the
bandwidth of the conduction band allowing an effective BCS Hamiltonian
to hold within an acceptable accuracy.
% 5) 
\item Only the Zener $s$-$d$ exchange with antiferromagnetic exchange
	amplitude $J_{sd}$ incorporated in the BCS approach gives an explanation of the well-known $T_{c,\,\mathrm{max}}$ -- Fermi contour shape correlation~\cite{Pavarini:01}.
% *) 
\item To the best of our knowledge
the first and so far the only theoretical explanation of this
correlation has been most likely solved.
%Due to lack of alternative explanation for more than 20 years we arrive at the conclusion that a long standing puzzle has been possibly solved.
%\end{itemize}
\end{enumerate}

%\subsection{Who's Afraid of Virginia Woolf? And what has to be put in the condensed matter theory agenda?}

\subsection{What is next on the condensed matter theory agenda?}
It is unlikely that the Kondo exchange $J_{sd}$ integral~\cite{Kondo} can be
calculated via first-principles.
However, the significant correlation coefficient $66\%$ depicted in
\Fref{Fig:tpt-lambda_} hints that for cuprates its value is weakly
material dependent, and its constancy is an acceptable initial
approximation.
We suppose also the same for the logarithm of
Euler-Mascheroni energy $\ln E_\mathrm{C}$ 
\Eqref{Euler-Asynchronism_energy_}.

Let us recall the physical meaning of the hybridization parameter
$\chi_\mathbf{p} \equiv S_\mathbf{p} D_\mathbf{p}$.
According to the electron band 
functions~\cite[Eq.~(1.4) and Eq.~(2.3)]{MishonovPenev:11}
of a plane wave with quasi-momentum 
$\mathbf{P}=\hbar\mathbf{p}/a_0$
\begin{align}
\Psi_\mathbf{p}(\mathbf{r})=
\frac{\exp(\mathrm{i}\mathbf{p}\cdot\mathbf{n})}
{\sqrt{N}}
\sum_{\mathbf{n}}
&\left[
S_\mathbf{p}\,
\psi_{\mathrm{Cu}4s}(\mathbf{r}-\mathbf{r}_\mathbf{n})
\right.
\label{LCAO_wave_function}
\\
&
+D_\mathbf{p}\,\psi_{\mathrm{Cu}3d_{x^2-y^2}}
(\mathbf{r}-\mathbf{r}_\mathbf{n})
\nn\\
&+X_\mathbf{p}\,\psi_{\mathrm{O}2p_{x}}(\mathbf{r}-\mathbf{r}_\mathbf{n}-\mathbf{R}_x)\mathrm{e}^{\mathrm{i}\varphi_x}
\nn\\
&\left. 
+Y_\mathbf{p}\,\psi_{\mathrm{O}2p_{y}}(\mathbf{r}-\mathbf{r}_\mathbf{n}-\mathbf{R}_y)\mathrm{e}^{\mathrm{i}\varphi_y}\right],\nonumber
\end{align}
where\begin{align}
&
\mathbf{R}_x=\frac12(1,\, 0) a_0,\qquad
\mathbf{R}_y=\frac12(0,\, 1)  a_0,\qquad
N=N_xN_y,\nn\\
&
\varphi_x=\frac12 (p_x-\pi),\qquad
\varphi_y=\frac12 (p_y-\pi),\qquad
\mathbf{n}=(n_x,\,n_y),
\nn\\
&
n_x=0,1,\dots, N_x\gg1,\qquad
n_y=0,1,\dots, N_y\gg1,\nn
\end{align}
the hybridization parameter 
is nothing but the probality an electron to be simultaneously in the
$\mathrm{Cu}4s$ and the $\mathrm{Cu}3d_{x^2-y^2}$.
In real space the Heisenberg-Fermi operators of annihilation
and creation obey the anti-commutator relations
\cite[Eq.~(7.3)]{LL9}
\be
\hat{\Psi}_\alpha(t,\mathbf{x})\hat\Psi_\beta^\dagger(t,\mathbf{y})
+\hat\Psi_\beta^\dagger(t,\mathbf{y})\hat\Psi_\alpha(t,\mathbf{x})
=\delta(\mathbf{x}-\mathbf{y})\delta_{\alpha,\beta},
\label{Fermi_anti_commutator}
\ee
while in \Eqref{LCAO_wave_function} 
$\Psi_\mathbf{p}(\mathbf{r})$ is the complex single particle
wave function.
The use of one and the same notation $\Psi$ for amplitude
and operators coined the terminology \textit{second quantization}.
The letters $S$ and $D$ comes from the standard $s$ and $d$
conventional notations for the orbital momentum values $l=0$ and
$l=2$, and the corresponding orbitals $\psi_{\mathrm{Cu}4s}(\mathbf{r}-\mathbf{r}_\mathbf{n})$
and
$\psi_{\mathrm{Cu}3d_{x^2-y^2}} 
(\mathbf{r}-\mathbf{r}_\mathbf{n})$
of a copper ion localized at a lattice point
$\mathbf{r}_\mathbf{n}$.
The notations $X_\mathbf{p}$ and $Y_\mathbf{p}$
come from the orientation of oxygen 2$p$ orbitals
$\psi_{\mathrm{O}2p_{x}}(\mathbf{r}-\mathbf{r}_\mathbf{n}-\mathbf{R}_x)$ and
$\psi_{\mathrm{O}2p_{y}}(\mathbf{r}-\mathbf{r}_\mathbf{n}-\mathbf{R}_y)$.
$\Psi_\mathbf{p}(\mathbf{r})$ is the one electron wave function
projected on the Cu$4s$, Cu$3d_{x^2-y^2}$,
O$2p_{x}$, and O$2p_{y}$ orbitals which are relevant to
the CuO$_2$ plane superconductivity.
In the 4-band second quantization Hamiltonian 
\Eqref{4_band_LCAO} the operators $\hat S_{\mathbf{n},\alpha}$,
$\hat D_{\mathbf{n},\alpha}$, $\hat X_{\mathbf{n},\alpha}$
and $\hat Y_{\mathbf{n},\alpha}$ are the corresponding
Fermi annihilation operators for an electron orbital in the $\mathbf{n}$ elementary cell of the CuO$_2$
plane.
Analogously in the Kondo interaction Hamiltonian
\Eqref{Eq.Kondo} $\hat S_{\mathbf{n},\alpha}^\dagger$,
$\hat D_{\mathbf{n},\alpha}^\dagger$ are the electron creation 
operators for the Cu4$s$ and the Cu$3d_{x^2-y^2}$ orbitals of the
Cu ion in the $\mathbf{n}$ elementary cell.
Finally, the
$\hat{S}_{\mathbf{n},\,\alpha}^\dagger
\hat{D}_{\mathbf{n},\,\beta}^\dagger
\hat{S}_{\mathbf{n},\,\beta}
\hat{D}_{\mathbf{n},\,\alpha}$
second quantization terms describe how these orbitals
exchange their electrons.

The complex one electron wave function \Eqref{LCAO_wave_function}
corresponds to a Bloch wave in the LCAO approximation.
If we wish to represent in real space
the Heisenberg Fermi annihilation operator electron field in the LCAO
representation, the corresponding operator field reads
\begin{align}
\hat\Psi_\alpha(\mathbf{r},t)=
\sum_{\mathbf{n}}
&\left[
\hat S_{\mathbf{n},\alpha}(t)\,
\psi_{\mathrm{Cu}4s}(\mathbf{r}-\mathbf{r}_\mathbf{n})
\right.
\label{LCAO_ahhihilaton_field}
\\
&
+\hat D_{\mathbf{n},\alpha}(t)\,\psi_{\mathrm{Cu}3d_{x^2-y^2}}
(\mathbf{r}-\mathbf{r}_\mathbf{n})
\nn\\
&+\hat X_{\mathbf{n},\alpha}(t)\,\psi_{\mathrm{O}2p_{x}}(\mathbf{r}-\mathbf{r}_\mathbf{n}-\mathbf{R}_x)
\nn\\
&\left. 
+\hat Y_{\mathbf{n},\alpha}(t)\,\psi_{\mathrm{O}2p_{y}}(\mathbf{r}-\mathbf{r}_\mathbf{n}-\mathbf{R}_y)\right].\nonumber
\end{align}
The 4-band LCAO Hamiltonian \Eqref{4_band_LCAO} contains
Fermi creation and annihilation operators with the anti-commutator
relations, for example
\be
\hat{S}_{\mathbf{n},\alpha}(t)\,\hat{S}_{\mathbf{m},\beta}^\dagger(t)
+\hat{S}_{\mathbf{m},\beta}^\dagger(t)\,\hat{S}_{\mathbf{n},\alpha}(t)
=\delta_{\mathbf{n},\mathbf{m}}\,\delta_{\alpha,\beta}.
\label{Fermi_anti_commutator_real_space}
\ee

In the momentum space, the square of the modulus the corresponding
Bloch wave amplitudes gives the probability
\begin{equation}
\label{chi2-SD}
\mathsf{P}_{sd} = \chi_\mathbf{p}^2
\equiv |S_\mathbf{p}D_\mathbf{p}|^2
\end{equation}
with
\begin{equation*}
|S_\mathbf{p}|^2+|D_\mathbf{p}|^2
+|X_\mathbf{p}|^2+|Y_\mathbf{p}|^2=1
\end{equation*}
of a band electron with momentum $\mathbf{p}$ to be simultaneously 
in the orbitals $\mathrm{Cu}4s$ and $\mathrm{Cu}3d_{x^2-y^2}$.
According to this interpretation 
the main ingredient of the cuprate pairing theory
\begin{align}
\overline{\mathsf{P}}_{\!\! sd}&=
\rho_\mathrm{_F}\langle\chi_\mathbf{p}^2\rangle
=\sum_b \int 
\mathsf{P}_{\!\! sd}\,
\delta(\varepsilon_{\mathrm{b},\,\mathbf{p}} - \varepsilon_{\mathrm{_F}})\,\frac{\mathrm{d}^D p}{(2\pi)^D}
\nn\\
&=\sum_b \int 
|S_{\mathbf{p},\mathrm{b}}D_{\mathbf{p},\mathrm{b}}|^2\,
\delta(\varepsilon_{\mathrm{b},\,\mathbf{p}} - \varepsilon_{\mathrm{_F}})\,\frac{\mathrm{d}^D p}{(2\pi)^D}
\end{align}
is the average of the $s$-$d$ hybridization probability 
on the Fermi surface and where it is necessary to perform 
summation on all conduction bands $b$ in the
$D$-dimensional $\mathbf{p}$-space.
The summation should be included in
the LCAO wave function \Eqref{LCAO_wave_function}.
Due to the $\delta$-function in the integrand,
$\overline{\mathsf{P}}_{sd}$ has dimension 1/energy.
Finally the dimensionless BCS coupling constant is given by
\be
\lambda_\mathrm{_{BCS}}
=2 J_{sd}\,\overline{\mathsf{P}}_{\!\! sd}.
\ee

The LCAO approach is a good approximation for the cuprate electron band structure.
It provides the unique possibility to reveal the relation between the averaged hybridization probability
$\langle\chi_\mathbf{p}^2\rangle$ and the dimensionless parameter 
$t^\prime/t$
determining the shape of the Fermi contour in the 2d approximation.
This is a current demonstration that physical intuition is the best scientific instrument.
The calculation of the electron band structure of many cuprates
deserved considerable efforts -- the analysis of the Fermi contour
shape was the crucial experiment clarification for the pairing
mechanism.
In spite of the fact that high-$T_c$ cuprates are midst the best studied materials,
one can easily claim that there is a need for more precise studies to
check the preliminary conclusions obtained via LCAO.
The $s$-$d$ interaction amplitudes 
$S_\mathbf{p}$ and  $D_\mathbf{p}$ can be calculated directly by
partial wave analysis on atomic spheres used in muffin-tin \textit{ab
initio} electron band calculations.

The Fermi surface averaging of the hybridization probability is then a
routine numerical task. Finally, the relation
$1/\ln T_c$ versus $\overline{\mathsf{P}}_{sd}$ has to be independently studied
by \textit{ab initio} band calculations.
The slope of this linear regression will determine the corresponding value of the
Kondo exchange integral $J_{sd}$. This task should be put on the agenda of
contemporary condensed matter physics.
In parallel the matrix elements of all other pairing interactions can
be easily evaluated.
Having some partial clarification, we feel obliged to speculate on
some questionable and disputable condensed matter problems.
If we are on the correct track, we have to obtain much more than we have invested -- an open question is whether the Kondo interaction can be incorporated in the 
Abrikosov~\cite{Abrikosov:03} scenario for
spin density waves (SDW) mechanism 
of the metal insulator transition in the parent undoped cuprates, 
perhaps as $s$-$d$ waves mechanism.
In brief, owing universality, we would like to check in what other processes
the possible \textit{s-d} pairing mechanism plays a primordial role.

In his study Abrikosov~\cite{Abrikosov:03} 
used our LCAO approach~\cite{MishonovPenev:00}, 
\cite[Chap.~1]{MishonovPenev:11}
for independent particles Hamiltonian.
To this Hamiltonian he adds an appropriate exchange interaction
which induces the metal-insulator transition in the parent material
or density waves in doped cuprates. We can analyze
some alternative variations of his
standing density waves (SDW) idea projected on the LCAO basis.
We suppose that super-exchange amplitudes
in this basis $J_{dd}$, $J_{sp}$, $J_{pd}$ and $J_{sd}$
can bring about a metal insulator transition at zero doping
and small pseudo-gap close to optimal doping.
In Ref.~\cite{OMahony:22} it was observed that 
the phase of pseudo-gap follows the phase of the apex modulation and again
$J_{sd}$ is suspicious for producing a pseudo-gap as well.
The implementation of \textit{s-d} interaction in the 
developed Abrikosov~\cite{Abrikosov:03} scheme
is a technical problem which can be included in the agenda of
condensed matter theory. Here, different scenarios are
possible:
1) Charge density waves by Cu$3d_{x^2-y^2}$
density at Cu ions $\langle
\hat D_{\mathbf{n},\sigma}^\dagger\hat D_{\mathbf{n},\sigma}
\rangle$ 
and related to this charge density waves.
2) SDM modulation only of the spin densities
$\langle
\hat D_{\mathbf{n},+}^\dagger\hat D_{\mathbf{n},+}
\rangle$ 
and $\langle
\hat D_{\mathbf{n},-}^\dagger\hat D_{\mathbf{n},-}
\rangle$
at zero charge modulation
$\langle
\hat D_{\mathbf{n},\sigma}^\dagger\hat D_{\mathbf{n},\sigma}
\rangle
=\langle
\hat D_{\mathbf{n},+}^\dagger\hat D_{\mathbf{n},+}
\rangle
+\langle\hat D_{\mathbf{n},-}^\dagger\hat D_{\mathbf{n},-}
\rangle $.
3) A hybridization modulation
$\langle
\hat S_{\mathbf{n},+}^\dagger\hat D_{\mathbf{n},-}\rangle+
\langle\hat S_{\mathbf{n},-}^\dagger\hat D_{\mathbf{n},+}
\rangle+\cdots$ for which self-consistent potential, in which all
electrons move, is created by the Kondo amplitude $J_{sd}$.
The anisotropy of the pseudegap is an important hint to a
dominant superexchange amplitude.
In the well-known ARPES picture adapted in the interesting 
review~\cite[Fig.~5]{Wang:20}, the Fermi contour is clearly seen at
the diagonals of the Brillouin zone $p_x=p_y$ and this suggests the
importance of the $S_p$ amplitude. Here, only the diagonal matrix
elements of the Kondo interaction have zeros.

Having a theoretical scheme to evaluate theoretically
the critical temperature of optimally doped cuprates, it is challenging 
to try to take a step further.
The filling of the Cu4s orbital significantly contributes
to the Knight shift of $^{63}$Cu. 
For a recent review on NMR of cuprates see the review 
by Haase~\cite{Haase:22} and references therein.
It will be usefull to perform a LCAO approximation of the Knight
shift
in an acceptable approximation even along the maximal gap directions
$S_\mathbf{p}^2\ll D_\mathbf{p}^2\simeq 1$.
This means that one can expect an additional correlation between
the Knight shift $K(T>T_c)$ and $\ln T_{c,\mathrm{max}}$.
It will be a significant hint in favor of the exchange mediated 
superconductivity.
Besides the $s$-$d$ Kondo exchange
the other exchange terms and even phonon attraction
will give a de-pairing contribution to $d$-wave superconductivity.
Their contribution can be revealed in the study of doping
dependence of the critical temperature $T_c(\tilde p)$
approximated by the parabola \Eqref{doping_dependence}.
The negative isotope effect in cuprates can also find a natural 
alternative explanation within this scheme, cf. Ref.~\cite{Abrikosov:00}.

The success associated to reaching high-$T_{c,\mathrm{max}}$ is based on a triple proximity 
of the Cu3$d_{x^2-y^2}$ level close to the O2$p$ level and the Cu4$s$ level not too far.
Analogously the triplet proximity of O2$p$, Cu3$d_{x^2-y^2}$ and
Cu4$s$ levels lies in the basis of suggesting the
Kondo $s$-$d$ as predominant pairing mechanism.

\subsection{Modulation experiments by Scanned Josephson Tunneling Microscopy~\cite{OMahony:22}}
\label{SJTM}

\subsubsection{How does the height of the CuO$_5$ pyramid change the gap?}

The significant correlation coefficient of the data represented in
\Fref{Fig:lmbTc_} hints that the \textit{s-d} mechanism deserves
further attention. This result is not a firm
proof to identify the correct pairing mechanism
even when this correlation is in the
limelight of high-$T_c$ physics for more than 20 years without
being backed by a robust theoretical explanation.
Moreover, the correlation coefficient of the linear regression shown on
FIG. \ref{Fig:lmbTc_} is not exactly unity as it is for the case of the
isotope effect for the correlation between the critical
temperature $T_c$ and the atomic mass $M_a$;
$\ln T_c$ versus $\ln M_a$ plot for a single metal.
In \Fref{Fig:lmbTc_} data are depicted for different CuO$_2$
compounds and for all of them the CuO$_2$ Hamiltonian is perturbed
by out of plane atoms.
Recently a \textit{subtle}~\cite{Lord:82} method for perturbing the
CuO$_2$ plane in a single compound was developed in Ref.~\cite{OMahony:22}.
The straight line in their figure ~\cite[Fig.~5C]{OMahony:22}
has a remarkable correlation coefficient
which deserves further theoretical analysis.

In electronics, for example, low voltage AC measurements 
are much more sensitive, accurate, and reliable than DC measurements.
Analogously the structural super-modulation in
Bi$_2$Sr$_2$CaCu$_2$O$_{8+x}$ studied by
Josephson Scanning Tunneling Spectroscopy (JSTM)
gave perhaps the best confirmation of R\"ohler's
idea~\cite{Roehler:00,Roehler:00a} 
concerning the importance of the hybridization of the Cu$4s$ state
with the CuO$_2$ plane.
Recall that, hybridization is strongly influenced by the apex oxygen.
Hence, the space supermodulation of the 
distance between copper ion and apex oxygen
is the best tool to modulate the superconductor's order parameter
to test the validity of the alternative theoretical explanations of
pairing in cuprates.

In recent JSTM studies ~\cite{OMahony:22}, it was reported
that the space $\mathbf{r}$ image of the modulation of
maximal Josephson current $I_{J}$ induced by structural modulation of
the distance $h$ of the apex oxygen with respect to CuO$_2$ plane.
The height $h(\mathbf{r})$ of the CuO$_5$ pyramid have up to
$2\iota=12\%$ modulation in the crystal $a$-direction
\begin{align}
&
h(x_a)=h_0[1+\iota_h\cos(Q_ax_a)],\quad Q_a=2\pi/L_a,
\label{h(x_a)}
\\
&
L_a\sim 26\,\text{\AA},\quad h_0= 2.44\,\text{\AA},
\quad \delta h\equiv\iota_h\, h_0 = 0.14\,\text{\AA},\nn \\
&
\iota_h = 0.0574.\nn
\end{align}
%\hl{Now we have to describe how the apex oxygen orbital}
%$O2p_z$ localized at 
%\be
%\mathbf{R}_a(\mathbf{n})=h_\mathbf{n}\mathbf{e}_z,\qquad 
%\mathbf{e}_z=(0,\,0,\,1).
%\ee
To the LCAO Hamiltonian $\hat H_\mathrm{LCAO}$  
\Eqref{4_band_LCAO}
we have to add the hybridization of apex $O2p_z$  oxygen with the Cu$4s$ orbital
\begin{align}
\hat H_\mathrm{apex}  =  -\sum_{\mathbf{n},\alpha} 
t_{sa}(\mathbf{n})
\left(
\hat{S}_{\mathbf{n},\alpha}^\dagger\hat{Z}_{\mathbf{n},\alpha}
+ 
\hat{Z}_{\mathbf{n},\alpha}^\dagger\hat{S}_{\mathbf{n},\alpha}
\right)
\label{apex}
\end{align}
and to the LCAO wave function \Eqref{LCAO_wave_function}
the contribution of all O$2p_z$ apex orbitals written
in coordinate space
\be
\sum_{\mathbf{n}} 
Z_\mathbf{n}\psi_{\mathrm{O}2p_{z}}(\mathbf{r}-\mathbf{r}_\mathbf{n}-\mathbf{R}_a(\mathbf{n})).
\ee
Due to the block matrix structure of the Hamiltonian
in every pyramid we have to solve a single particle  Hamiltonian
\be
H_\mathrm{apex} =\begin{pmatrix} \epsilon_s^{(0)}& -t_{sa}\\
-t_{sa}&\epsilon_p
\end{pmatrix},
\label{band_Hamiltonian}
\ee
which slightly correct the position of the Cu$4s$ level,
cf~\cite[Eq.~(2.62)]{Migdal} and 
\cite[Sec.~39, Problem 1, Eq.~(79.4)]{LL3}
\begin{align}
 \epsilon_s^{(0)}\rightarrow  \epsilon_s
& =\frac{ \epsilon_s^{(0)}+\epsilon_p}{2}
 +\sqrt{\left(\frac{ \epsilon_s^{(0)}+\epsilon_p}{2}\right)^2+t_{sa}^2 }
 \nn\\
 &
\approx \epsilon_s^{(0)} +\Delta \epsilon_s,
\qquad
\Delta \epsilon_s\equiv
\frac{t_{sa}^2}{\epsilon_s^{(0)}-\epsilon_p},
\label{Delta_epsilon_s}\\
\chi_{sa}&\equiv SZ\approx 
\frac{t_{sa}}{\epsilon_s^{(0)}-\epsilon_p}\ll 1,\nn
\end{align}
for $t_{sa}\ll \epsilon_s^{(0)}-\epsilon_p$.
%%% New version starts here -- 25 May 2025
Analogously
\be
\epsilon_{p,z}=\frac{\epsilon_s^{(0)}+\epsilon_{p,z}^{(0)}}2
-\sqrt{\left(\frac{\epsilon_s^{(0)}-\epsilon_{p,z}^{(0)}}2\right)^2
+t_{as}^2}
<\epsilon_{p,z}^{(0)}.
\label{epsilon_pz}\\
\ee
%The apex Hamiltonian \eqref{apex}
The band Hamiltonian in momentum representation
\Eqref{band_Hamiltonian}
is diagonalized by a Jacobi rotation in 
the $S_\mathbf{p}$-$Z_\mathbf{p}$ plane
\begin{subequations}
\begin{align}
&
\begin{pmatrix}
c_\varphi&s_\varphi\\
-s_\varphi&c_\varphi
\end{pmatrix}
\begin{pmatrix}
\epsilon_s^{(0)} &t_{as}\\
t_{as} &\epsilon_{p,z}^{(0)}
\end{pmatrix}
\begin{pmatrix}
c_\varphi&-s_\varphi\\
s_\varphi&c_\varphi
\end{pmatrix}
=\begin{pmatrix}
\epsilon_s&0\\
0&\epsilon_{p,z}
\end{pmatrix},\\
&
\begin{pmatrix}
S_\mathbf{p}^\prime\\
Z_\mathbf{p}^\prime
\end{pmatrix}
=
\begin{pmatrix}
c_\varphi&s_\varphi\\
-s_\varphi&c_\varphi
\end{pmatrix}
\begin{pmatrix}
S_\mathbf{p}\\
Z_\mathbf{p}
\end{pmatrix},\\
&
\begin{pmatrix}
S_\mathbf{p}\\
Z_\mathbf{p}
\end{pmatrix}
=
\begin{pmatrix}
c_\varphi&-s_\varphi\\
s_\varphi&c_\varphi
\end{pmatrix}
\begin{pmatrix}
S_\mathbf{p}^\prime\\
Z_\mathbf{p}^\prime
\end{pmatrix},
\label{Reverse_Jacobi}
\end{align}
\end{subequations}
where
\begin{subequations}
\begin{align}
%&
%c_\varphi=\cos\varphi_a,\quad s_\varphi=\sin\varphi_a,\quad\\
&
\tan(2\varphi_a)=\frac{2t_{as}}
{\epsilon_s^{(0)}-\epsilon_{p,z}^{(0)}}>0,\\
&
s_\varphi=\sin\varphi_a=\frac1{\sqrt{2}}\sqrt{1-\dfrac{1}
{\sqrt{1+\left[\tan(2\varphi_a)\right]^2}}},\\
&
c_\varphi=\cos\varphi_a=\frac1{\sqrt{2}}\sqrt{1+\dfrac{1}
{\sqrt{1+\left(\dfrac{2t_{as}}{\epsilon_s^{(0)}-\epsilon_{p,z}^{(0)}}\right)^2}}}.
\label{cos_phi_apex}
\end{align}
\end{subequations}
This approximate diagonalization 
 actually gives a narrow band.
As the apex O$2_{p,z}$ strongly hybridizes with the
Cu$4s$ the main amplitude of the tunnel current
is strongly influenced by the apex distance
and the energy parameter
\be \mathcal{E}(h)\equiv\epsilon_d-\epsilon_{p,z}.
\ee
As it is emphasized in 
Ref.~\cite[Fig.~5C]{OMahony:22},
the simultaneous observation of 
super-modulation of $\mathcal{E}(h)$ and
super-fluid density $n_p(h)$ gives 
\emph{the long sought} tool to revile the pairing mechanism.

For a double layer $\mathrm{(CuO_2)_2}$, we have analogously
$\delta \epsilon_s=-t_{ss}$, where $t_{ss}$ is the transfer amplitude between
two close Cu$4s$ orbitals. 
These perturbations are crucial to our understanding of
the pressure dependence on $T_c$.
For model evaluations, we can use the Harrison~\cite{Harrison:80} 
interpolation formula
\be
t_{sa}(h)=\eta_{sp\sigma}\frac{\hbar^2}{m_e h^2},\qquad
\eta_{sp\sigma}\simeq1,
\label{t_sa_Harrison}
\ee
which explains how modulations of the pyramid height $\delta h$
affects the position of Cu$4s$ level.
The subscripts of the Harrison dimensionless parameter
$\eta_{sp\sigma}$ mean that we have a $\sigma$-valence bond
between an $s$ and a $p$ orbital, as it is for the $t_{sp}$ transfer amplitude.
The interatomic dependence $\propto 1/h^2$ leads to
\be
t_{sa}(h)=\left(\frac{a_0}{2h}\right)^{\!2}t_{sp} = 0.61 \, t_{sp}. %=1.21~\mathrm{eV}.
\label{t_sa_t_sp}
\ee
In order to unveil the influence on the height $h$ of the pyramid
on the superconducting gap $\Delta_\mathbf{p}=\tilde\Xi(T)\tilde\chi_p$
parameterized by the order parameter $\Xi$ and 
the gap anisotropy function $\tilde\chi_p$,
we differentiate the zero temperature order
parameter $\tilde\Xi_0\equiv\tilde\Xi(0)$ with respect to $h$
using the chain rule
\be
\frac{\md \tilde\Xi_0}{\md h}=
\frac{\md t_{sa}}{\md h} \cdot
\frac{\md \epsilon_s}{\md t_{sa}} \cdot
\frac{\md \mathcal{F}}{\md \epsilon_s} \cdot
\frac{\md T_c}{\md \mathcal{F}} \cdot
\frac{\md  \tilde\Xi_0}{\md T_c},
\label{Quality_height}
\ee
where ``$\cdot$'' is used as a delimiter.
Let us describe the beginning of the chain:
From \Eqref{t_sa_Harrison} and \Eqref{t_sa_t_sp}
we have
\be
\frac{\md t_{sa}}{\md h}
=-\frac{2 t_{sa}}{h_0},\quad t_{sa}=t_{sa}(h_0),
\ee
and the subscript ``$_0$'' will be further omitted.
The second order perturbation formula
\Eqref{Delta_epsilon_s} gives
\be
\frac{\md \epsilon_s}{\md t_{sa}} 
= \frac{2t_{sa}}{\epsilon_s^{(0)}-\epsilon_p}.
\ee
For the last derivative multiplier, we have the standard BCS ratio.
Now, we differentiate expression \eqref{t'/t_}
\be
\mathcal F=\mathcal{A}/(2\mathcal{A}+4\mathcal{B}),
\ee
with respect to the position of the Cu$4s$ level $\epsilon_s$.
A straightforward differentiation yields
\begin{align}
\frac{\md \mathcal{F}}{\md \epsilon_s} 
&=\mathcal{F}\,\frac{\dot{\mathcal A }}{\mathcal{A}}
-\mathcal{F}^2
\frac{2\,\dot{\mathcal A}+4\,\dot{\mathcal B}}{\mathcal{A}},
\label{dot_F}
\\
\dot{\mathcal A}&\equiv 32\,t_{pp}
\left(t_{pd}^2+t_{pp}\varepsilon_d\right),\\
\dot{\mathcal B}&\equiv 4\,\varepsilon_pt_{pd}^2,
\end{align}
where the energy is taken at the Fermi level
$\epsilon=\eF$.
As $t_{pp}$ is a smaller, almost negligible parameter,
we have $\dot{\mathcal A}\ll\dot{\mathcal B}$,
then from \Eqref{TcBCS_}, using a differentiation
and the numerical value of $\lambda$ from
Table~\ref{tbl:out_energy_}
and a graphical differentiation from \Fref{Fig:tpt-lambda_}
allows to calculate another ``sub-chain'' of the product of
derivatives
\begin{align*}
\frac{\md T_c}{\md \mathcal{F}}
&=\frac{\md T_c}{\md \lambda^{-1}}\cdot
\frac{\md \lambda^{-1}}{\md \lambda}\cdot
\frac{\md \lambda}{\md \mathcal F}, \\
\frac{\md T_c}{\md \lambda^{-1}}&=-T_c,\qquad
\frac{\md \lambda^{-1}}{\md \lambda}=-\frac1{\lambda^2},\qquad
\frac{\md \lambda}{\md \mathcal F}&=0.588,
\nn\\
\frac{\md T_c}{\md \mathcal{F}}&=0.588\,\frac{T_c}{\lambda^2}.
\end{align*}
Finally, \Eqref{pi_gamma}, \cite[Eq.~(40.4)]{LL9}
and \cite[Eq.~(34.23)]{AbrGorDzya}
give the last multiplier in the chain \label{First_chain}
\begin{subequations}
\begin{align}
\frac{\md T_c}{\md  \tilde\Xi_0}&
=\frac{\gamma}{\pi}
= 0.56693295955\dots\approx 0.57,\\
\frac{\md  \tilde\Xi_0}{\md T_c}&
=\frac{\tilde\Xi_0}{T_c}
=1.76387\dots,
\end{align}
\end{subequations}
and we can calculate the derivative of the order parameter
with respect to the height of the pyramid
\begin{align}
&
F_h\equiv\frac{\md \tilde\Xi_0}{\md h}=|q_e|E_h,\qquad
E_h\approx 5.16\,\text{mV/\AA},
\end{align}
which has the dimension of force; the notation 
$q_e=1.602\,\mathrm{C}$ is the modulus of electron charge.
The large value of the so defined electric field parameter $E_h$
motivates us to introduce the dimensionless logarithmic derivative 
\begin{subequations}
\label{Quality_Xi^2}
\begin{align}
Q_{\Xi^2}
 & \equiv\frac{\md\ln\Xi^2}{\md \ln h}
 =2\frac{h}{\Xi_0}\frac{\md \Xi_0}{\md h} =1.855 \cos^2(\varphi_a)
	=1.759 \\
& =-4 \, \frac{t_{sa}}{t_{sp}} \cdot \frac{2 t_{sa}}{\epsilon_s^{(0)}-\epsilon_p} \cdot \Gamma_\mathrm{RA} \cdot \frac{0.588}{\lambda^2} \cos^2(\varphi_a)
\end{align}
\end{subequations}
and analyze after some rearrangement the  physical meaning of the
separate multipliers. The additional factor $\cos^2\varphi_{a}$ arises because
the Jacobi rotation according to \Eqref{chi2-SD}
slightly changes $S_\mathbf{p}$ which gives the substitution 
$\chi^2\rightarrow\chi^2\cos^2\varphi_{a}$.
For comparison with the experiment, it is more convenient to use
dimensionless parameters, which somehow unifies the ideas 
by R\"ohler (R) for the importance of the Cu$4s$ level and
the Andersen (A) Fermi contour form factor 
$\mathcal F \equiv t^\prime/t$ arises
\be
\Gamma_\mathrm{RA}\equiv t_{sp}\frac{\md \mathcal{F}}{\md\epsilon_s}
=-9.3 \times 10^{-3}.
%=-6.00\times 10^{-3}.
\ee
For this parameter the LCAO approach
provides the analytical expressions \Eqref{dot_F}.
%%% End new version in purple color
%% Old version here
There are no especially small parameters and 
$Q_h$ have an undetectable large value.
As the effect is proportional to $t_{as}$ the influence of apex oxygen
on $T_x$ almost vanishes if Cu4$s$ orbital is removed from the LCAO analysis.

Now we can explore how the modulations of the apical distance $h$
and the superconducting gap observed by the
Josephson effect are related.
%For a nice review of physics of pair density waves
%see for example Ref.~\cite{Wang:20}.

\subsubsection{Recalling the formulas for the Josephson effect}
Let us recall the Ambegaokar-Baratov formula connecting the maximal Josephson current $I_J$ and the conductivity $G=1/R_N$
of the contact in the normal state between two 
isotropic gap superconductors with different gaps, for a detailed derivation see the book by 
Abrikosov~\cite[Eq.~(22.14)]{Abrikosov}
\begin{subequations}
\begin{align}
q_e I_J R_N&=2\frac{\Delta_1\Delta_2}{\Delta_1+\Delta_2}
\mathrm{K}((\Delta_1-\Delta_2)/(\Delta_1+\Delta_2)),\\
\mathrm{K}(k)&\equiv\int_0^{\pi/2}\frac{\md \varphi}
{\sqrt{1-k^2\sin^2\varphi}},
\end{align}
\end{subequations}
for $\Delta_1,\,\Delta_2\ll T_c$.
The theoretical idea of this calculation
is very simple.
The conductivity of the normal phase $1/R_N$
is proportional to the averaged on both the Fermi surfaces transfer amplitudes between
the two superconductors
$\langle\langle 
t_{1,2}^2
\rangle\rangle_{_{1,2}}$.
The averaged matrix elements are in turn proportional to
the maximal Josephson current $I_J$.
These properties of the Josephson contact are reliably canceled 
in the ratio $U_J\equiv R_NI_J$ which depends only on the properties of bulk superconductors in both sides of the tunneling transition.

The tunneling matrix elements propotional to $\chi_\mathbf{p}\chi_{sa}$
have a structural supermodulation. 
This self supermodulation is canceled in the ratio
of the maximal current $I_j$ and the conductivity $R_N^{-1}$
in the Josephson contact.
It seems that in SJTM, the distance tip to sample can be tuned to
maintain the normal conductivity $1/R_N$ constant.

For anisotropic gap superconductors, we have some complications
related to orientations, etc, 
but for a qualitative analysis, we use 
the order parameter $\Xi$ which parameterizes the superconducting gap.
Let both gaps be almost equal
\be
\Xi_1=\Xi_0,\qquad 
\Xi_2=(1+\iota)\,\Xi_0,\qquad
|\iota|\ll1.
\ee
Using $\mathrm{K}(0)=\pi/2$
for the Josephson voltage 
\be
U_J\equiv I_J R_N,
\ee
the Ambegaokar-Baratov formula yields
\be
U_J=\frac{\pi}{2}\frac{\Xi_0}{q_e}\left(1+\frac{\iota}2\right).
\ee
As for small $\iota\ll1$ when $\Delta_1\approx\Delta_2$ we may use (
cf. Ref.~\cite{Kimura:09})
\begin{subequations}
\begin{align}
&
\dfrac2{\dfrac1{1+\iota}+1}\approx\left(1+\frac{\iota}2\right)
\approx\sqrt{1+\iota},\\
&
\frac{2\Delta_1\Delta_2}{\Delta_1+\Delta_2}\approx\sqrt{\Delta_1\Delta_2}.
\end{align}
\end{subequations}
The Feynman two level picture~\cite[Eq.~(21.44)]{FLP3}
gives the same result and one can consider
that $\Xi_1$ and $\Xi_2$ are wave functions $\Psi$ of the superconducting condensate and superconductivity
is quantum mechanics in classical aspect.

For anisotropic gap superconductors, the formula for the Josephson
voltage has an additional factor
\be
U_J\equiv I_J R_N=\frac{\pi}{2}\frac{\overline\Xi_0}{q_e}
|\cos\alpha_{12}|,
\ee
where $\alpha_{12}$ is the angle between the normalized gap 
anisotropies when the tip is rotated to an angle $\varphi$
with respect to the superconductor surface
\begin{subequations}
\begin{align}
\cos\alpha_{12}&\equiv \langle\langle 
\overline{\chi}_{\mathbf{p}_1}\overline{\chi}_{\mathbf{p}_2}
\rangle\rangle_{_{1,2}},\\
\overline{\chi}_{\mathbf{p}}&=
\frac{\chi_{\mathbf{p}}}
{\sqrt{\langle
\chi_{\mathbf{p}}^2}\rangle}\approx\cos(l\varphi), \qquad
\Delta_\mathbf{p}=\overline{\Xi}(T)\overline{\chi}_{\mathbf{p}}.
\end{align}
\end{subequations}
For d-wave CuO$_2$ superconductors ($l=2$)
the Josephson voltage has a cusp like minimum at
$\phi=45^\circ$, while for SrRuO$_4$ it is at
$\phi=90^\circ$.
It would be useful if SJTM can reveal what is the gap anisotropy 
in SrRuO$_4$, to this end the samples should be put in different orientations 
for one and the same tip of the tunneling microscope.

\subsubsection{Modulation of Cooper pair density and Josephson current}

The most direct way to study the influence of the 
structural supermodulation
is to simultaneously inspect the behaviors of the height of the pyramids
\eqref{h(x_a)}
and the %introduced in Ref~\cite{OMahony:22} 
normalized electron pair density defined
by the square of space averaged Josephson voltage 
$U_J^2$ \cite{OMahony:22} via
\begin{align*}
&
n_p(\mathbf{r})\equiv|\Psi|^2\propto [U_J(\mathbf{r})]^2,
\qquad
\overline{n}_p(\mathbf{r})\equiv\frac{n_p(\mathbf{r})}
{\left<n_p(\mathbf{r})\right>_\mathbf{r}},\\
&
n_p(x_a)=n_{p,0}[1+\iota_{n}\cos(Q_ax_a)).
\end{align*}
%See the experimental dependences \cite[Fig.~5B]{OMahony:22}.
The ratio of the modulation depths is a measure to evaluate
the effect of the structural supermodulation on the superconducting
order parameter
\be
Q_n\approx\frac{\iota_{n}}{\iota_h}.
\ee
Another estimation of the same parameter yields the logarithmic derivative
extracted through the linear regression of the experimental data 
\cite[Fig.~5C]{OMahony:22}
\be
Q_n=\left. \frac{h}{\overline{n}_p}
\frac{\md \overline{n}_p}{\md h} 
\right|_{h_\mathrm{max}} \approx 1.965.
\ee
The comparison between the \emph{s-d}-LCAO
theory and the experiment is given by the ratio of the theoretically calculated logarithmic derivative
without fitting the parameters
\eqref{Quality_Xi^2} and the experimental one above
\be
\frac{Q_n}{Q_{\Xi^2}}=\frac{1.965}{1.759} \approx 1.12.
\ee
Initially, the theory must have only a qualitative agreement with the
experiment, i.e. a correct sign and the same order of magnitude of the
effect. To conclude, O'Mahony et al~\cite{OMahony:22} are correct that
their SJTM modulation of the distance between planar Cu and apical O
atom in optimally doped Bi$_2$Sr$_2$CaCu$_2$O$_{8+x}$
($T_c=91$~K and $\tilde p=0.17 \pm0.01$)
reveals the pairing mechanism of copper-oxide high-$T_c$ superconductors.

Finally, we discuss the sign meaning. The lowering of the height
of the pyramids promotes an increase of the hybridization of the
apex O$2p_z$ the plane Cu$4s$.
This leads in turn to a perturbative increase of the energy of the Cu$4s$ level 
and lowering of the Cu$4s$ and Cu$3d$ hybridization.
The intensity of the \textit{s-d} interaction decreases,
and this finally leads to a critical temperature decrease.

Scanned data for the Josephson voltage $U_J(x,y)$
and the apex oxygen distance $h(x,y)$
can be depicted in one plot: $I_JR_N$ in ordinate and 
the pyramid height in the abscissa.
Advanced experimental data processing requires
only a linear regression and the calculation
of the parameters of the straight fit by a least square line
$\langle U_J\rangle(h)$.
The slope of this line provides an important
parameter
\be
Q_h^\mathrm{(exp)}\equiv\frac{\md\langle U_J\rangle}{\md h}.
\ee
for SJTM microscopy.

\subsubsection{Josephson effect and fluctuation phenomena}
The strong modulation of the superconducting gap has an important
influence on the fluctuation the magnetic susceptibility
$\chi_{ab}$
of Bi$_2$Sr$_2$CaCu$_2$O$_{8+x}$ when the magnetic field is
perpendicular to the CuO$_2$ planes.

For every layered superconductor the in-plane fluctuation of the conductivity
$\sigma_{ab}$ is proportional to the susceptibility, so
that their ratio is temperature independent
\be
\tau_0=\frac{\mu_0}{3}\xi_{ab}^2(0)\frac{\sigma_{ab}}
{-\chi_{ab}}=\mathrm{const},\qquad \mu_0=4\pi\times10^{-7}
\ee
and parameterizes the lifetime constant
$\tau_0$ of the fluctuation of Cooper 
pairs~\cite[Eqs.~(5.59-65)]{MishonovPenev:11},
where the coherent length $\xi_{ab}^2(0)$ is determined
via temperature derivative of the upper critical field $B_{c2}$
normal to the layers~\cite[Eqs.~(5.15)]{MishonovPenev:11}
\begin{align}
-T_c\frac{\md B_{c2}(T)}{\md T}=\frac{\Phi}{2\pi\xi_{ab}^2(0)},
\quad \Phi_0=\frac{2\pi\hbar}{|e^*|},\quad
e^*=2e, 
\end{align}
and $e=-1.62\,10^{-19}$ is the electron charge.
The lifetime of the fluctuation of the Cooper pair which enters in the 
associated Boltzmann equation has a critical slowing-down
\begin{align}
\tau(T)=\frac{T_c\tau_0}{T-T_c},\;\;
\tau_0=\tau_0^\mathrm{(BCS)}\tilde{\tau}_\mathrm{rel},\;\,
\tau_0^\mathrm{(BCS)}=\frac{\pi}{16}\frac{\hbar}{T_c}.
\end{align}
The most famous example is the Aslamasov-Larkin conductivity
for 2d superconductors~\cite[Eqs.~(5.15)]{MishonovPenev:11}
\be
\sigma_\mathrm{AL}(T)=\frac{e^2}{16\hbar}\,
\tilde{\tau}_\mathrm{rel}\,\frac{T_c}{T-T_c}
=\frac{e^2T}{\pi\hbar^2}\,\tau(T).
\ee
However, samples demonstrating a strong modulation 
of the order parameter at low temperatures $T\ll T_c$
must exhibit also a strong modulation of the critical temperature. 
Whence, the fluctuation conductivity should be quasi-one dimensional
along the strips of the structural supermodulation and
simultaneously the fluctuation magnetic moment should be 
significantly suppressed because the fluctuation Cooper pairs
cannot rotate in an area of the order of 
$2\pi\xi_{ab}^2(0)/(T-T_c)=\Phi_0/B_{c2}(T)$.
For these samples one can predict that
the dimensionless relative lifetime
must have large values $\tilde{\tau}_\mathrm{rel}>1$.
While for Hg substitution crystals where the supermodulation
vanishes $\tilde{\tau}_\mathrm{rel}\approx 1$ and this
BCS value is irrelevant to the pairing modulation.

We were unable to find in the literature (see
Ref.~\cite{OMahony:22} and references therein) a Hamiltonian depending
on the apex distance.
The relevant to the considered SJTM experiment is the perturbation term
\be
\hat V_h\equiv h\,\frac{\partial\hat H}{\partial h},
\ee
which is useful for the calculation of the
temperature dependence of the gap $\Xi(R)$
and its height dependence at zero temperature 
$\Xi_0(h)$.
%In the next subsection we consider in short the used in
%Ref.~\cite{OMahony:22} formulae and notions.

\subsubsection{Notions and notations unified}
We follow almost all standard notations for the 
non-phonon magnetic (exchange) mechanisms of superconductivity.
For example, Anderson's theory is based on the \emph{t-J} model
with antiferromagnetic interaction $J_{dd}$ between neighboring Cu spins.
One alternative model is the spin-fluctuation or paramagnon
theory there are relatively simple diagrams showing how spin
fluctuations can cause pairing, see for example
\cite{Scalapino:99,Monthoux:07}.
Let us juxtapose the notations used by
O'Mahony et al~\cite[Fig.~1, Eq.~(1)]{OMahony:22}
and Pavarini et al~\cite{Pavarini:01}, 
following the notation used throughout the present study
\begin{align}
\mathcal{E} & =0.9\mathrm{~eV},\quad
t \equiv t_{pd},\quad e^2\equiv q_e^2/4\pi\varepsilon_0,
\nn \\
J&\equiv J_{dd}=\frac{4t^4}{\mathcal{E}^3}
=\frac{4t_{pd}^4}{\mathcal{E}^3}\ll U,
\label{super_exchange_J_dd}
\\
U_{ab}&\equiv \int\int \md^3 x\, \psi_a^2(\mathbf{x})\,
\frac{e^2}{|\mathbf{x}-\mathbf{y}|}\,
\psi_b^2(\mathbf{y})\,\md^3 y,\\
U&\equiv U_{dd}>U_{ss}\gg t_{sp}>t_{pd}\gg t_{pp}. \nn
\end{align}
Here we do not discuss the details of the calculation
of the super-exchange amplitude of two electrons in adjacent
Cu positions in the quadratic CuO$_2$ lattice.
The exchange takes place between the
Cu$3d_{x^2-y^2}$ and Cu$3d_{x^2-y^2}$ orbitals for
copper ions in the elementary cells $\mathbf{n}=(n_x,\, n_y)$
and $\mathbf{n}=(n_{x+1},\, n_y)$, for example.
As a side comment, we point out that there is no
physical hopping in $t^\prime$ along the diagonals in a quadratic
lattice between $(n_x,\, n_y)$ and ($n_{x+1},\, n_{y+1}$) lattice points.
This is just a trivial mathematical exercise for a tight binding model in a quadratic lattice
but not a real process in the CuO$_2$ plane.

For two electrons $J_{dd}$ is the super-exchange amplitude 
between the adjacent Cu ions, but what happens if the terminal points are Cu$4s$ and Cu$3d_{x^2-y^2}$?
In this case we simply have to replace $t_{pd}\rightarrow t_{sp}$.
Hence, super-exchange amplitude in 
\Eqref{super_exchange_J_dd} would read
\be
J_{sd}\approx4\,\frac{4t_{sp}^2 t_{pd}^2}{\mathcal{E}^3},
\qquad \mathcal{E}\approx \epsilon_d-\epsilon_p,
\ee
where the pre-factor 4 comes from the number of the oxygen ligands
around every Cu ion in the CuO$_2$ lattice.
The used parameters in Ref.~\cite{Pavarini:01} yield
\be
\frac{J_{sd}}{J_{dd}}\simeq 4\left(\frac{t_{sp}}{t_{pd}}\right)^2 \approx 7.
\ee
Formulas of this kind are derived within 
the four band (Cu$4s$, Cu$3d_{x^2-y^2}$, O$2p_x$, O$2p_y$)
Hubbard model.
In this sense, we completely agree with O'Mahony et al~\cite{OMahony:22} that the
conclusion that superexchange is the key to the
electron pairing mechanism of hole-doped cuprates.
Our result for the depth of modulation ratios
\Eqref{Quality_height} does not depend on the numerical value 
of $J_{sd}$.
These logarithmic derivatives depend solely on the superexchange
amplitude and the involved orbitals.
We only wish to express the hints that the larger amplitude
$J_{sd}>J_{dd}$
emanating from the number of ligands
has far better chance %in the Darwin selection principle
for Cooper pairing.
The other superexchange amplitudes $J_{pd}$ and $J_{pp}$
favor pairings with incorrect anisotropy and there could be only de-pairing terms describing the parabolic dependence 
\Eqref{doping_dependence}, an idea that is beyond the scope of this
study.
In any case, we emphasize all theoretical formulas
for the superexchange amplitudes reported above
\Eqref{super_exchange_J_dd} explain only qualitatively the 
philosophy of their derivation and are expressed by 
other unknown parameters. 
Even the simplest possible case of the
Heitler-London exchange of two electrons in a hydrogen 
%quasi-molecule
molecule H$_2$ when atoms are far 
$R\gg a_\mathrm{B}$
from their equilibrium distance 
\be
J_{ss}(R)\simeq
0.82 \frac{\hbar^2}{m_ea_\mathrm{B}^2}\,
(R/a_\mathrm{B})^{5/2}\exp(-2R/a_\mathrm{B})
\ee
is a sophisticated quantum 
problem~\cite[Sec.~81. Valency]{LL3}.
Even the simplest case of a single electron hopping in the H$_2^+$ ion 
at large distances between the protons $R\gg a_\mathrm{B}$
\be
t_{ss}(R)\simeq 2 \frac{\hbar^2}{m_ea_\mathrm{B}^2}\,
(R/a_\mathrm{B})\exp(-2R/a_\mathrm{B}-1),
\ee
where $m_e$ is the electron mass and $a_\mathrm{B}$ is the 
Bohr radius, can not be exactly reproduced by numerical methods
using electronic band calculations.

\subsubsection{Additional consideration for ARPES, LDA, SJTM and $T_c$}
Since their discovery, many high-$T_c$ cuprates were synthesized over a short time.
It became obvious that pairing is mediated by a quite omnipresent
interaction and cannot be based on an exotic subtle process.
On the other hand, all interactions in condensed matter physics are well-known and well described in contemporary textbooks and reviews.
Most of them (if not all) were attempted for the explanation of the high-$T_c$ mechanism.
The CuO$_2$ superconductors became by far the most studied materials,
but it what was the crucial experiment for our understanding of the pairing mechanism?

Our attention was attracted by the correlation
of $T_c$ and the Fermi contour form factor parameter
$\mathcal{F}=t^\prime/t$. The Fermi contour is observable by ARPES 
and can be reliably calculated within LDA.
Only the Kondo \textit{s-d} exchange interaction can explain
in a natural way the linear correlation $-\ln T_c$ versus the Fermi contour shape parameter $\mathcal{F}$.
How can the $J_{sd}$ exchange survive in different condensed matter areas?
The anisotropy of the scattering rate in the normal phase 
qualitatively follows the gap anisotropy in the superconductor phase
and simultaneously the $T$-linear dependence of the in-plane resistivity.
A much stronger argument is however the support of SJTM where the
\textit{s-d} interaction even quantitatively explains the modulation
depth.
%Thermodynamics, $T_c$, kinetics of the Ohmic resistance, 
%LDA, ARPES and SJTM are the \hl{distribution} areas where $J_{sd}$ fits excellently.
Thus, we are witnessing the emergence of a coherent picture
with no necessary exotic parameters to be included.
%Of course, there are other interactions influencing this picture, but
%the one with the controlling stake has been found.
Other exchange interactions can be relevant to spin waves
\cite{spin_waves}, yet they can be only a small
de-pairing perturbation for the superconducting order
parameter.

\subsection{Normal state conductivity}
\label{Sec: Normal state conductivity}
Having at hand a coherent picture for pairing in the CuO$_2$ plane,
we may use it at least qualitatively in the 
whole range of phenomena related to electronic processes 
in CuO$_2$ plane superconductors.
Even a qualitative explanation of this phenomenon suggests how
important is the Hamiltonian leading to pairing at low temperatures.

The phase diagram of the high-$T_c$ cuprate
superconductor Pb-doped Bi$_2$Sr$_2$CaCu$_2$ O$_{8+\delta}$ was
revisited by anisotropic transport measurements in Ref. \cite{Harada:22}.
It was found that the conductivity anisotropy parallel and perpendicular to the CuO$_2$ planes varies from approximately 25 for YBCO$_7$ 
to 550 for optimally doped Bi(Pb):2212.
This strong anisotropy provides strong grounds to use as first approximation
2d models explaining superconductivity in a single CuO$_2$ plane.
Whence, for ARPES data we often use the expression ``Fermi
contour'' instead of ``Fermi surface'';
CuO$_2$ planes are almost insulated and the normal transport between
them above $T_c$ is as a rule related to incoherent hopping.
Thus, perpendicularly to the planes, electron transport cannot
be related to the pairing mechanism, rather it is determined by 
out-of-plane structures. Let us note that, a single superconducting 2d plane have
a gapless plasmon spectrum with frequency proportional to the square
root of the wave-vector $\omega\propto \sqrt{k}$
\cite{MishonovGroshev:94,Buisson:95}.

As it was predicted a long time ago~\cite{Mishonov:91},
Bi$_2$Sr$_2$CaCu$_2$ O$_{8+\delta}$ is transparent in the far
infrared, because the plasma frequency $\Omega_z$ for the electric
polarization perpendicular to the CuO$_2$ double layers is below the
maximal value of the superconducting gap $\Delta$, i.e
$\hbar\Omega_z<\Delta$.

In the normal phase $T>\Delta$ the weak incoherent transport between 
the (CuO$_2$)$_2$ layers opens up new degrees of freedom
related to the fluctuations of the electric field perpendicular to the layers.
We consider the electric field $E_z$
fluctuations in this layered (strange) metal.
Applying this idea to infinite CuO$_2$ layers,
we have to take into account space inhomogenities
of the perpendicular electric field $E_z(x,y)$.
As the plasma frequency is smaller than the temperature
$\Omega_z<T$, these fluctuations are purely classical.
%Then it is possible to apply the equipartition theorem.
%Using the fact that the fluctuations of the electric field $E_z(x,y)$ are 
%electrostatically related to the 2D density of the charge carriers of the
%conducting planes of the layered metal, the classical equipartition theorem can be applied to the fluctuations of the 2D charge density of conducting layers.
Owing to the equipartition theorem, the fluctuations of 2D density in the conducting layers indispensably leads to 
proportional to the temperature in-plane electrical resistivity
$\varrho_{ab}\propto T$.
Qualitatively this scattering remains the Rayleigh scattering
of the sunlight by density fluctuations of the atmosphere
which leads to the blue color of the sky.

For superconductors with moderate anisotropy where
$\hbar\Omega_z\simeq T_c$, the
electric field fluctuations are almost frozen and 
the electron-electron interaction leads to
the standard Fermi liquid behavior
$\varrho_{ab}=\varrho_0+A_{_\textrm{BLP}} T^2$ (
see the monograph by Mott~\cite{Mott:90}
Baber~\cite{Baber:37} and 
Landau-Pomeranchuck~\cite{Landau:36}), then
both processes can be detected in layered cuprates.

The matrix elements of the electron-electron interaction 
have different angular dependence leading to an anisotropy of the charge carriers lifetime
$\tau_\mathbf{p}$ along the Fermi contour.
Moreover, the Coulomb electron-electron repulsion induces a negligible
anisotropy of the lifetime~\cite{Mishonov:00}, while \textit{s-d}
exchange interaction in the Born approximation gives for the scattering
rate~\cite{hotspot2022}
\be
\frac1{\tau_\mathbf{p}}\propto T\,J_{sd}\,\chi_\mathbf{p}^2\, .
\label{life_time}
\ee
This behavior is in a qualitative agreement with the 
anisotropy obtained by ARPES measurements.
Thus, at least qualitatively
$T_{c,\mathrm{max}}$ versus $t^\prime/t$,
SJTM of the supermodulation and ARPES data for the 
anisotropy of width of the ARPES lines are in a qualitative
agreement. 
%One important interaction must have dominant influence in 
%completely different phenomena.

Finally, we consider qualitatively the in-plane normal conductivity 
of CuO$_2$ layers.
%The concept of mean free path 
%$l=v_\mathrm{F}\tau$ was introduced by Clausius before
%Boltzmann to create science kinetics.
We will recall the Drude formula for the in-plane conductivity 
\be
\sigma=\frac{ne^2}{m}\langle \tau_\mathbf{p} \rangle,
\ee
where $\langle \tau_\mathbf{p} \rangle$ is the lifetime averaged along  the Fermi contour.
The conductivity of parallel connected resistors is given by
individual conductivities $1/R_i$, i.e.
\be
\frac{1}{R}=\frac{1}{R_1}+\frac{1}{R_2}+\frac{1}{R_3}+\cdots\,.
\ee
Yet, according to \Eqref{life_time}, the scattering intensity via
exchange interaction vanishes along the diagonals of the
Brillouin zone $|p_x|=|p_y|$.
%In this could spots of the Fermi contour scattering is determined
%by another terms of electron-electron interaction, 
%most probably the omnipresent Coulomb repulsion.
%In this case, one could spots of the Fermi contour make shortcut
%of the hot spots where hybridization and superconducting gap in the low temperatures have maximum.
Thus, averaging of the lifetime 
$\langle \tau_\mathbf{p} \rangle$
along the Fermi contour indispensably will include other
processes that are not essential to the pairing of optimally doped cuprates
and its critical temperature $T_{c,\mathrm{max}}$. 

When the scattering rate is much smaller than the Fermi energy
$\hbar/\tau\ll\epsilon_{_F}-\epsilon_d$ and
the mean free path is much larger than the lattice constant
$l\gg a_0$, we can approximate the charge carriers by a gas medium.
Furthermore, even for the coherence length 
$\xi_0\ll l$, we have a clean superconductor whose effective mass of
the super-fluid charge carriers is approximately
ten times the effective mass of normal carriers. 
In brief the inequalities for a gas medium
\be
\frac{\hbar/\tau_{\mathbf{p},\mathrm{max}}}
{\epsilon_{_F}-\epsilon_d},\;
\frac{a_0}{l},\; R_w-1 \ll1
\label{normal_state_approximations}
\ee
explain the success of BCS approach for the explanation
of the studied correlation between $T_{c,\mathrm{max}}$
and the shape of the Fermi contour.
Finally, the conductivity in the perpendicular to the
CuO$_2$ plane is mainly incoherent 
and depends on the specific ions between the conducting layers.
Moreover, it is often a strongly doping dependent.

%%%%%%%%%%%%%%%%%%%%%%%%%%%
\section*{Acknowledgments}
The authors wish to acknowledge TMM’s former
collaborators in this long endeavor: 
Ted Sariyski in 1980-1981 organized in Sofia
a seminar on high-$T_c$ superconductivity
in which represented the idea that $s$-$d$
exchange can be a mechanism for high-$T_c$.
Discussions with Nikolay Tonchev
and Jordan Brankov on the \textit{s-d}
pairing
mechanism %~\cite{Tonchev:80,Bogolyubov:1981,chamati1992a,chamati1994b}
is highly appreciated.
In 2002 in Berlin solving BCS equations for many exchange 
interactions in CuO$_2$ plane 
Evgeni Penev observed $d_{x^2-y^2}$
symmetry for the superconducting gap for
Kondo $s$-$d$ interaction.
Later on, together with Zlatan~Dimitrov,
Stanislav~Varbev, Karin~Omar, 
Alexander~Stefanov and Evgeni~Penev was
performed the preliminary study~\cite{BJP:11} of
$T_{c,\mathrm{max}}$ versus Fermi contour
correlation, described shortly in the 
textbook~\cite{MishonovPenev:11}.
The interest to this study, comment and suggestions by
\'{A}lvaro~de~R\'{u}jula, Valery~Pokrovsky, 
Stefan Drechsler, Damyan~Damyanov, Mauro Doria, 
Valya Mishonova, Hermann Suderow, 
Michael Speckmann,
Abdou Hassanien, Vladimir Fomin and
Andrey~Varlamov are highly appreciated.
One of the authors (TMM) appreciate 
the discussion long time ago
with Gilbert Lonzarich how Kondo interaction
in singlet channel creates attraction leading to pairing.
The initial version of the present study was represented
in the conference devoted to 90-th birthday of Valery Pokrovsky.

The authors acknowledge the support by Grants K$\Pi$-06-H58/1 dated
15.11.2021 and K$\Pi$-06-H78/2 dated 05.12.2023 of the Bulgarian National Science Fund
and the European Regional Development Fund under
``Research Innovation and Digitalization for Smart
Transformation'' program 2021-2027 under the Project
BG16RFPR002-1.014-0006 -- National Centre of Excellence in
Mechatronics and Clean Technologies.
Authors are also thankful to their friends Atanas Batinkov, Yoana Ruseva and Nikolay Aleksandrov for the interest and support of this work.
The main ingredient for successive work of the
\textit{s-d} exchange as a pairing mechanism
is based on the unique proximity of 3 perovskite levels
O$2p$, Cu$3d$, and Cu$4s$. 
This approximate energy coincidence remains
the formula for the Spring vacation:
we have to wait for equinox, then full moon and
finally for weekend; happy Vacation.

\section*{Author contributions}
All authors have equally contributed to the writing of the manuscript, programming, making of figures  and experimental data processing.\\

\section*{Data availability statement}
The data that support the findings of this study are 
available after arXiv version 4 of the current manuscript
\href{https://arxiv.org/abs/2303.18235}{arXiv:2303.18235 [cond-mat.supr-con]}.
It should be noted that EISPACK part of the LAPACK package E.~Anderson~\textit{et~al},
\textit{LAPACK Users' Guide}, 3-rd ed. (Philadelphia~PA, Soc. Indust. and Appl. Math., 1999)  for solution of the eigenvalue problem and Matplotlib library 
J.~D.~Hunter, Matplotlib: A 2D graphics environment,
\href{https://doi.org/10.1109/MCSE.2007.55}{IEEE Comp. Soc.~\textbf{9}(3), pp.~90-95 (2007)}
for producing graphs have been used in this study.
% The source file of the manuscript together with FORTRAN
% programs of the figures is published in the arXiv version 
% of the present 
% study~arXiv:2303.18235 [cond-mat.supr-con]
 %\cite{arXiv_Pokrovsky}.

%\bibliography{Pokrovsky}

\end{document}